\documentstyle[preprint,prb,eqsecnum,aps]{revtex}

\begin{document}
\draft
\preprint{}
\title{Spin-Charge Separation in the $t-J$ Model:\\ Magnetic and Transport
Anomalies}
\author{ Z.Y. Weng, D.N. Sheng, and C.S. Ting }
\address{Texas Center for Superconductivity and Department of Physics\\
University of Houston, Houston, TX 77204-5506 }
\maketitle
\begin{abstract}
A real spin-charge separation scheme is found based on a saddle-point state of
the $t-J$ model.  In one-dimensional (1D) case, such a saddle-point reproduces
the correct asymptotic correlations at the strong-coupling fixed-point of the
model. In two-dimensional (2D) case, the transverse gauge field confining
spinon and holon is shown to be gapped at {\em finite doping} so that a
spin-charge deconfinement is obtained for its first time in 2D. The gap in the
gauge fluctuation disappears at half-filling limit, where
a long-range antiferromagnetic order is recovered at zero temperature and
spinons become confined.  The most interesting features of spin dynamics and
transport are exhibited at finite doping where exotic {\em residual} couplings
between spin and charge degrees of freedom lead to systematic anomalies with
regard to a Fermi-liquid system. In spin dynamics,
a commensurate antiferromagnetic fluctuation with a small, doping-dependent
energy scale is found, which is characterized in momentum space by a Gaussian
peak at ($\pi/a$, $ \pi/a$) with a doping-dependent width ($\propto
\sqrt{\delta}$, $\delta$ is the doping concentration). This commensurate
magnetic fluctuation contributes a non-Korringa behavior for NMR spin-lattice
relaxation rate. There also exits a characteristic temperature scale below
which a pseudo-gap behavior appears in the spin dynamics.
Furthermore, an incommensurate magnetic fluctuation is also obtained at a {\em
finite} energy regime. In transport, a strong short-range phase-interference
leads to an effective holon Lagrangian which can give rise to a series of
interesting phenomena including linear-$T$ resistivity and $T^2$
Hall-angle. We
discuss the striking similarities of these theoretical features with those
found in the high-$T_c$ cuprates and
give a consistent picture for the latter. Electronic properties like
Fermi-surface and superconducting pairing in this framework are also discussed.

\end{abstract}

\vspace{0.4in}

\narrowtext

\newpage

\section{INTRODUCTION}

The  normal state of  high-$T_c$ cuprate superconductors  has shown
peculiar properties in both charge and spin channels.  In transport aspect, the
resistivity\cite{res} exhibits a linear-temperature dependence up to $1000 $
$K$ and down to a temperature $\simeq T_c$, which can be as low as $10$ $K$.
This temperature dependence has been related to a scattering rate\cite{opt}
that behaves like $\eta \frac{k_B}{\hbar} T$ with $\eta \sim 2$ for all  the
optimally doped cuprates with $T_c$ ranging
from 10 up to over 100 $K$ (a linear-frequency dependence of the scattering
rate at frequency $>k_BT/{\hbar}$ is also found in the infrared
spectroscopy\cite{opt} up to $0.15$ $eV$). Such a linear-$T$ longitudinal
resistivity is also
accompanied by a Hall coefficient \cite{hall} which implies hole-like charge
carriers and shows a $1/T$ dependence in contrast to a $T$-independent Fermi
liquid case. A   Hall-angle experimental analysis\cite{angle1,angle2} has
demonstrated that the
$1/T$ behavior in the Hall coefficient is due to an additional scattering rate
in the transverse channel, which behaves like $T^2$. Most recently, the
magneto-resistance has been
found to have  a $T^{-4}$ temperature dependence,\cite{mr} at variance with the
Kohler's rule.  Furthermore, the thermopower has shown  a strong
doping-dependence,\cite{thermo}
which decreases with the increase of doping and even changes sign in the
overdoped regime. All of these transport results are anomalous with regard to
the canonical phenomena in a conventional Fermi liquid (FL) system.

Spin magnetic properies  have also exhibited  a number of  anomalies, which
persist into  the superconducting phases. Two powerful probes of spin dynamics
in the cuprates are nuclear-magnetic-resonance\cite{nmr} (NMR) and
neutron-scattering.\cite{neutron} In the cuprates,  NMR spin-lattice relaxation
rate $T_1^{-1}$, which probes a very small energy scale ($\sim 10^{-4}$ $meV$),
has shown a strong enhancement\cite{nmr} at low temperature for  planar  $Cu$
nuclei, in contrast to the usual  Korringa behavior in a Fermi liquid. On the
other hand, $T_1^{-1}$ for planar $O$ nuclei\cite{nmr} is more or less close to
the Korringa law.   The sharp contrast of the NMR $T_1^{-1}$ between the planar
$Cu$ and $O$ nuclcei strongly suggests\cite{wals,mila,shastry,hammel,mmp,bulut}
that the non-Korringa behavior at $Cu$ sites should be caused by
antiferromagnetic (AF) correlations among the $Cu$ spins whose effect could be
effectively cancelled out at $O$ sites.
AF spin fluctuations  have been also directly measured by the inelastic
neutron-scattering experiments\cite{neutron} at higher energies in
$La_{2-x}[Sr, Ba]_xCuO_4$ ($LSCO$) [Ref.\onlinecite{lsco,hayden,inc}]
and $YBa_2Cu_3O_{7-\delta}$ ($YBCO$) [Ref.\onlinecite{ybco1,ybco2,mook}]
materials. However, in contrast with more or less universal behaviors in NMR
spin relaxation rates,
neutron-scattering data have shown distinctive characteristics among these
compounds. For example,  commensurate AF fluctuations
are observed in the insulating $LSCO$  systems\cite{lsco,hayden} and underdoped
$YBCO$  compounds\cite{ybco1,ybco2}
with  characteristic energy scales much smaller than the exchange energy
$J\simeq 120$ $ meV$. But incommensurate AF correlations have been
found\cite{inc} in the metallic $LSCO$ down to an energy scale as low as
$1$ $meV$. Nevertheless, recent analyses have revealed\cite{walstedt,barzykin}
that the NMR relaxation rates in $LSCO$ are
inconsistent with an incommensurate AF structure at $\omega \simeq 0$. This is
because the effect of incommensurate fluctuations could not be well cancelled
out at $O$ sites, and a {\em large} non-Korringa signal would leak to the
latter, as in contradiction to experiment. In the optimally-doped $YBCO$, the
{\em absence} of low-energy AF
fluctuation in neutron-scattering experiment\cite{mook} is also inconsistent
with the NMR measurements.\cite{nmr} Thus a consistent explanation of the NMR
data and a reconciliation of NMR and neutron-scattering are two essential
issues in spin dynamics.

Given these anomalous transport and magnetic properties,  one
would naturally  question if  a single type
of elementary excitations carrying {\em both}  charge and spin can explain
them without leading to any intrinsic contradiction. Since the sharp
Fermi edge has been well identified in the cuprates by the angle-resolved
photo-emission measurements,\cite{olson,dessau} such
low-lying electron-like excitations, if exist,  should always locate near
the  Fermi surface,  no matter whether it is a Fermi liquid or a marginal
Fermi liquid (MFL) [Ref.\onlinecite{mfl}] system. The main difference between
FL and MFL lies in
their energy spaces (the broadening of the quasi-particle in MFL is linear
in energy\cite{mfl} as compared to $\omega^2$ in FL case). But in the phase
space (momentum space), they are facing essentially the same
problems.\cite{littlewood} Low-lying
states as labelled by momenta are uniformly distributed in momentum space
around Fermi surface. The only conceivable structure would come  from the
topology of Fermi surface. The question is whether such a structure of
Fermi surface (maybe with  nesting and van Hove singularity) is capable of
explaining an army of abundant experimental anomalies. In transport channel,
one of the key problems is how to get hole-type charge carriers, since the
elementary excitation is of electron-type.
Even though a special curvature of Fermi surface may explain the
hole-like sign indicated by the Hall effect, the total charge carrier number
could by no means coincide  with the hole number as
measured with regard to the half-filled insulating parent compound. Many other
problems could arise within such a framework. For
example, a monotonic decrease of the thermopower\cite{thermo} from over
$100$$\mu V/K$  down
to $0$ and even changing sign, when  doping is increased from zero to an
overdoped level, is  difficult to  understand here because the phase-space is
too limited. Even with the presence of a van Hove singularity, an overall
change as estimated by theory ($\sim 8$$\mu V/K$ [Ref.\onlinecite{vh}]) is too
small as compared to the experimental value. Without additional structure,  a
second scattering rate implied\cite{angle1,angle2} in the
Hall-angle measurements is even more difficult to comprehend, especially the
different ways by which  it gets into the transverse and longitudinal
channels.

In the spin channel, a number of FL and MFL theories\cite{fl,littlewood} have
been  proposed in terms of the geometry of  Fermi surface. The  topology of
Fermi surface determines essential characteristics of spin dynamics through the
Lindhard-type response function. These theories provide an explanation of
incommensurate AF
structure\cite{inc} in the metallic $LSCO$, but fail to account for other
equally important issues. One is about the NMR relaxation rates whose
aforementioned distinctive behaviors at planar $Cu$ and $O$ nuclei are
difficult to reconcile.\cite{fl,littlewood} The fundamental reason is that a
Lindhard response function cannot produce a sharp enough AF peak near ($\pi/a$,
$\pi/a$). Furthermore, the overall width of the AF peak in these theories is
generally not
sensitive to doping concentration, and one cannot find NMR anomalies at low
temperature to be sufficiently  enhanced with the decrease of doping as
indicated by the experiments.\cite{nmr}  Most seriously, the itinerant
picture fails to provide a small AF energy scale, as the
only characteristic scale is the  Fermi energy
$\epsilon_F$.  Although a peak at small energy may be introduced by van Hove
singularity,\cite{fl} it is hard to be related to a magnetic
energy. The exchange energy scale  $\sim J$ as indicated in various magnetic
measurements, and an even smaller doping-dependent magnetic energy scale found
by inelastic neutron-scattering \cite{lsco,hayden,inc,ybco1,ybco2} is a strong
indication that the local spin description
of AF fluctuation may be  more appropriate than an itinerant picture. This is
further supported by the high temperature $T_1^{-1}$ measurement in $LSCO$
[Ref.\onlinecite{imai}]
which shows that doping effect is basically evaporated when $T>600$$K$, where
$T_1^{-1}$ at finite doping coincides with that of an insulating
antiferromagnet, described by the localized spins under Heisenberg
model.\cite{sokol}

Therefore, the phase-space limitation in an itinerant description, which
becomes inevitable in a system where fermionic elementary excitations carry
both spin and charge, prevents a consistent explanation of rich phenomena in
the high-$T_c$ cuprates. This implies that an electron in the cuprates may be a
composite particle, consisting of more elementary excitations which carry
spin and charge quanta (spinon and holon) separately. In this way,  spinon
and holon may find enough phase-spaces of their own  to account for
experimental anomalies. A   spin-charge
separation scenario for the cuprates has been first  proposed by
Anderson.\cite{anderson,kivelson}  Under this idea,
a strong on-site Coloumb repulsion can lead to a split of energy band with a
large Hubbard gap, and thus real charge carriers (holons) in the conduction
band  will have equal number as that of doped holes controlled by experiments.
It is a significant first step  towards understanding the transport properties.
However, in order to  reconcile with a sharp Fermi edge found by the
angle-resolved photo-emission, a fermionic
spinon with a similar Fermi surface has to be assumed.\cite{anderson} Thus at
least in the spin channel, basic properties should not be very different from
conventional FL or MFL theory,
which is a serious set back. Later it has been further
realized\cite{baskaran,larkin,lee,ioffe} that there actually exists a
strong coupling among spinons and holons, known as gauge interaction,
and such a gauge force plays a role essentially to confine\cite{fradkin} spinon
and honon together, like the quark confinement\cite{polyakov} in quantum
chromodynamics. In other words,  whether there is a real spin-charge separation
in a 2D strongly-correlated model is still unclear.

So we are in a paradox situation.  Experimental measurements, on one hand,
point to a possibility of spin-charge separation.  Strong-correlation theories,
on the other hand, lead to spin-charge confinement. This awkward situation
is actually caused by an oversimplified choice to let spinons to have a large
Fermi surface, which undermines a local description for spins. One may
argue that in order to recover a correct electron Fermi surface--an
experimental constraint--a large spinon Fermi surface would be  necessary. But
this is not the case even in one dimension, where a real spin-charge
separation
exists in the Hubbard model and an electron Fermi surface (points) still
satisfies\cite{ogata} the Luttinger volume theorem. Even though usually people
may be used to the thinking\cite{and1} that a spinon Fermi surface is the
reason
for an  electron Fermi surface in 1D, the various correlations have never been
correctly derived under such a picture. In fact, a FL description of spinons is
a too rough approximation in 1D. In the strong-coupling (large $U$) regime, a
correct spin-charge separation  description\cite{weng1} has been established in
a path-integral formalism,\cite{weng3} where an electron is described as a
composite particle of a spinon
and a holon, {\em together} with a nonlocal phase-shift field. It is this
phase-shift field that  helps to recover the right Fermi surface position.  In
this formalism, spinons are described by a local spin representation
without a Fermi surface, and both spin and density, as well as various
electronic correlation functions have been correctly obtained.\cite{weng1} An
important lesson learned here is that the electronic Fermi surface is no longer
essential in a spin-charge separation scheme, and
it can be reproduced from a {\em phase-shift}  in a correct spin and charge
separation scheme. (We note that in 1D an another powerful method is the
bosonization method\cite{haldane,frahm,schulz,ren} in weak-coupling regime.
There the non-Fermi-liquid like low-lying spin and charge processes can be
directly described  near the {\em electronic} Fermi points. But its possible
generalization to two dimensions has many difficulties and is still under
investigation.)

This leads to a new 2D spin-charge scenario for the  cuprates. In this
scenario, spinons would be free of ``duty'' to be responsible for electron
Fermi surface, so that  they could get  sufficient phase-space freedom  to
describe anomalous spin dynamics. Electronic Fermi surface would be produced by
extra phase-shift field associated with the  fermion-statistics of electron,
and should  not be
directly related to holon and spinon.  This  provides a fundamental reason
for  the  Luttinger theorem to  be valid even in strong correlations.  It
also means a Fermi-surface topology is no longer  crucial in determining
spin and charge dynamics. That is,  the shape of a Fermi surface due to the
detailed band structure, which may vary from one material to others in the
cuprates,  should not be so relevant to the basic spin and charge anomalies in
a spin-charge  separation framework. Therefore it may well
justify   using a  simple one-band $t-J$ model\cite{anderson,zhang,rice} to
describe the  realistic cuprates if a spin-charge separation is indeed present
there. It is  noted that
several interesting experimental features have been  already  known for the
$t-J$ model. For example,  spin degree of freedom is described  in  a local
spin representation, which reduces to the Heisenberg Hamiltonian at
half-filling and well describes\cite{heisenberg,aa} the magnetic behaviors in
the insulating cuprates. On the other hand,  charge carriers  in this system
are naturally holes as measured from  the half-filled insulating phase, which
is also  a strong indication of the
experimental relevance for the model. A spin-charge separation is already
well-known for such a model in 1D case.

It is the purpose of the present paper to establish the above-described
spin-charge separation scheme within the $t-J$ framework. Like other
approximations,  we cannot directly prove that this 2D spin-charge separation
state is
{\em the} solution of the $t-J$ model. But such a  state  will satisfy the
following important criteria. The transverse gauge fluctuation will be found
suppressed in the long-wavelength and low-energy regime so that spinon and
holon are indeed deconfined at
finite doping (i.e., spin-charge separation). This is in contrast to singular
infrared gauge fluctuation\cite{larkin,lee,ioffe}
in the case of uniform RVB state. At zero doping limit, the gap of  the gauge
fluctuation will vanish
such that the spinons become confined again to form spin-wave excitations.  In
this case, a  long-range AF order will emerge at zero temperature. Furthermore,
the present state can also reproduce the correct results in 1D case, which
is an important check of the theory because only in 1D one has exact
solution.\cite{lieb}

In the present paper, the effective Hamiltonian describing the spin-charge
separation saddle-point will be derived in the following  form:
\begin{equation}
H_{eff}=H_s+H_h,
\end{equation}
where $H_s$ and $H_h$ are the spinon and holon Hamiltonians, respectively,
defined by
\begin{equation}
H_s=-J_s \sum _{<ij > \sigma } \left( e^{i\sigma A_{ij}^h}\right) b_{i \sigma
}^+b_{j\sigma } + h.c. ,
\end{equation}
\begin{equation}
H_h=-t_h\sum_{<ij>}\left(e^{i[-\phi_{ij}^0+ A_{ij}^s]}\right) h_i^+h_j + h.c.
\end{equation}
Here $b_{i\sigma}$ and $h_i$ are called spinon and holon annihilation
operators,
respectively, which are connected to electron operator in Eq.(1.6)
below. $H_s$ and $H_h$ in Eqs.(1.2) and (1.3) would resemble a standard
tight-binding Hamiltonian if there are no phase fields $A_{ij}^h$,
$\phi_{ij}^0$ and $A_{ij}^s$ ($<ij>$  refers to two nearest-neighboring
sites on the lattice). Here $\phi_{ij}^0$ [as defined in Eq.(2.45)] represents
a $\pi$-flux threading through each plaquette. A tight-binding model under
$\phi_{ij}^0$ can be easily diagonalized.\cite{affleck} The
most prominent feature in Eqs.(1.2) and (1.3) is the presence of
$A_{ij}^s$ and
$A_{ij}^h$, the so-called topological phases.   $A_{ij}^h$ is defined by
\begin{equation}
A_{ij}^{h }=\frac 1 2  \\ \sum_{l \neq i,j} \mbox{Im ln} \left[\frac{z_i -
z_l}{z_j-z_l}\right] \\n_l^h  ,
\end{equation}
with complex coordinate $z_i=x_i+iy_i$ in 2D, and $n_l^h=h_l^+h_l$ as the holon
number operator. $A_{ij}^s$ is given by
\begin{equation}
A_{ij}^s =\frac 1 2 \sum_{l \neq {i,j} }\\ \mbox{Im ln} \left[\frac{z_i -
z_l}{z_j-
z_l}\right] \\ \left( \sum _{\sigma } \sigma n_{l \sigma} ^b \right) ,
\end{equation}
with $n_{l\sigma}^b=b_{l\sigma}^+b_{l\sigma}$ as the spinon number operator.

The usual  gauge coupling, which otherwise would confine spinon and holon
together, will be shown suppressed and has been neglected in Eq.(1.1).  Thus
the  residual  interaction between spinon and holon are solely mediated
through
$A_{ij}^s$ and  $A_{ij}^h$ in Eqs.(1.2) and  (1.3). In 1D case, the complex
coordinate $z_i$ is reduced to an 1D variable, and it is easy to show that
$A_{ij}^h$ and  $A_{ij}^s$  vanish in Eqs.(1.4) and (1.5).  Thus
spinon and holon are {\em decoupled} in 1D, and behave just like
free-particles
on their own tight-binding lattices. In 2D case, however, $A_{ij}^h$ and
$A_{ij}^s$
can no longer be gauged away. $A_{ij}^h$ in Eq.(1.4) describes fictitious
$\pi$-flux
quanta attached to holons  which are seen only by  {\em spinons} in $H_s$.
Hence $A_{ij}^h$  will play the role to introduce doping effect into spin
degree of freedom. On the other hand, $A_{ij}^s$ represents $\pi$-flux quanta
bound to spinons which can be only seen by {\em holons}. $A_{ij}^s$ will then
play the  role of a scattering source in holon transport.  So in the
spin-charge separation scheme, new types of scatterings are present, and we
will show in this paper that it is due to these unconventional forces that
magnetic  and transport anomalies are produced in consistence with those found
in  the high-$T_c$ cuprates.

Finally, let us outline how an electron is composed of the elementary
excitations, holon and spinon, in this scheme. For both 1D and 2D, the
electron
annihilation operator $c_{i\sigma}$ will be rewritten as
\begin{equation}
c_{i\sigma}= h^+_ib_{i\sigma}\left[ e^{-\frac{i}{2}\sum_{l\neq i}\theta_i(l)
\left(\sigma n_l^h-\sum_{\alpha}\alpha
n_{l\alpha}^b+1\right)}(-\sigma)^i\right],    \end{equation}
where $\theta_i(l)$ is defined in Eqs.(2.19) and (2.20).
Decomposition (1.6) can be understood as that annihilating an electron is
equivalent to creating a holon and annihilating a spinon, and, at the same
time, inducing overall nonlocal phase-shifts [in the brackets of Eq.(1.6)].
There are several ways to interpret
the  involvement of phase shift fields in Eq.(1.6).  Since the nonlocal fields
appearing in Eq.(1.6) resemble the Jordan-Wigner transformation in both 1D and
2D cases,\cite{jw,jw2d}  one may interpret them as statistical transmutations.
Here $h_i^+$ is
a hard-core bosonic operator,  and $b_{i\sigma}$ also satisfies bosonic
commutation relation for the same spin index $\sigma$ but anticommutation
relation for opposite spins [for details see Sec. II].  Thus these nonlocal
fields in Eq.(1.6) are to guarantee $c_{i\sigma}$  obeying an electronic
commutation relation. A deeper physics behind this  is related to the
phase-shift
idea,  whose important role in a strongly-correlated system  has been first
realized by Anderson.\cite{anderson,and1,ren} In 1D it has been explicitly
shown\cite{weng2} that an overall adjustment  of the system occurs when a hole
is doped into a large-$U$ Hubbard chain, in order to retain the Marshall sign
rule\cite{marshall} in the spin degree of freedom which is  decoupled from the
charge degree of freedom. This adjustment is found to be just represented by a
phase-shift shown in Eq.(1.6), and  it can lead to the
correct Luttinger liquid behavior of the single electron  Green's function.
Furthermore, a phase-shift in 1D can also be interpreted as reflecting the fact
that each holon carries a spin domain wall [see Ref.\onlinecite{weng1} and
Sec.III]. Here one important distinction from Anderson's original
phase-shift idea is that the many-body phase-shift in 1D will not only give
rise to the right non-Fermi liquid behavior, but also {\em shift} the Fermi
surface to
the position satisfying the Luttinger theorem. In other words, there is no need
to assume a right Fermi surface at the very beginning. In the present scheme,
the form of Eq.(1.6) can ensure a large electron Fermi surface for both 1D and
2D. But we shall discuss these electronic properties in a
separated paper for the sake of clarity. In that same paper, we also show that
when  holons
and  spinons are both Bose-condensed, the phase-shift field in Eq.(1.6) will
lead to a long-range {\em pairing} correlation of electrons (i.e.,
off-diagonal long-range order), and determine the
symmetry of the superconducting order parameter. In the present paper, we
mainly focus on the spin-charge separation formalism and explore the
corresponding spin dynamics and transport  properties.

The remainder of the paper will be  organized as the following. In section II,
we construct the above-described spin-charge separation scheme\cite{fb3} based
on the $t-J$ model. At half-filling limit, a long-range AF order can be
recovered in 2D, while the well-known spin-spin correlation is also
obtained in 1D.
At finite doping, the gauge fluctuation is shown to be suppressed so that one
has a real spin-charge separation. In section III, the spin dynamics in this
scheme is studied at finite doping. In 1D, the correct spin-correlation is
to be reproduced. In 2D, a number of interesting properties are discovered
which provide a consistent picture for the anomalous spin properties in the
cuprates. In section IV, the transport properties is investigated.  We
show that there exists an interesting scattering mechanism within the present
framework which leads an effective long-wavelength, low-energy Lagrangian. Such
a Lagrangian can give rise
to a series of exotic transport properties, in excellent agreement with the
experimental measurements. Finally, a conclusive summery will be given in
section V.

\section{ SPIN-CHARGE SEPARATION: FORMALISM}

In this section, a mathematical formalism of the  spin-charge separation state
will be  constructed based on  the $t-J$ model.

\subsection{The slave-boson formalism and mean-field states}

According to  Anderson,\cite{anderson} Zhang and Rice,\cite{zhang} low-energy
physics in the cuprates layers may be properly described by a single band $t-J$
 model. The t-J Hamiltonian  reads\cite{rice}
\begin{equation}
H_{t-J}=H_t+H_J  ,
\end{equation}
with a hopping term
\begin{equation}
H_t=-t \sum_{<ij>\sigma} c_{i\sigma}^+ c_{j\sigma}+H.c.   , \end{equation}
and a superexchange term
\begin{equation}H_J=J\sum_{<ij>}\left({\bf S}_i\cdot {\bf S}_j-\frac {n_in_j}
4 \right),     \end{equation}
in which ${\bf S}_i=\frac 1 2 \sum _{\sigma}   c_{i\sigma}^+(\hat{\bf
\sigma})_{\sigma \sigma '} c_{i\sigma '}$ (  $\hat{\bf \sigma}$ is the Pauli
matrix), and $n_i=\sum_{\sigma }  c_{i\sigma}^+ c_{i\sigma}$.  The Hilbert
space of the Hamiltonian Eq.(2.1) is
restricted by the no-double-occupancy constraint
\begin{equation}
\sum _{\sigma }  c_{i\sigma}^+ c_{i\sigma}  \leq 1,
\end{equation}
which imposes a strong correlation on electrons.

By using the slave-boson decomposition of the electron
operator\cite{barnes,zou}
\begin{equation}      c_{i \sigma } =h_i^+ f_{i \sigma },      \end{equation}
 the no-double-occupancy constraint (2.4) can be  replaced by an
equality constraint
\begin{equation}
h_i^+h_i+\sum _{\sigma }f_{i\sigma}^+f_{i\sigma}=1, \end{equation}
where $h_i ^+ $ is a bosonic  creation operator  and $f_{i \sigma } $ is a
fermionic annihilation operator, known as holon and spinon operators,
respectively. In this formulation, the hopping and exchange terms
of the $t-J$ Hamiltonian become
\begin{equation}
H_t=-t \sum_{<ij>\sigma} h_j^+h_i f_{i\sigma}^+ f_{j\sigma}+H.c.,
\end{equation}
and
\begin{equation}
H_J=-\frac J 2 \sum_{<ij> \sigma , \sigma ' }
  f_{i\sigma}^+ f_{j\sigma }f_{j\sigma '}^+ f_{i\sigma '}.      \end{equation}
In obtaining Eq.(2.8), a term $
\propto J\delta^2$ ($ \delta$ is the doping concentration ) has been
neglected\cite{lee} as usual for simplicity.

The advantage of using the slave-particle formulation for the $t-J$ model is
that once the initial state satisfies the constraint (2.6), the
no-double-occupancy
condition can be always preserved under the Hamiltonian (2.7) and (2.8).
Thus the constraint (2.6) in the slave-particle representation will not play
a crucial role as the constraint (2.4) does in the original Hamiltonians
(2.2) and (2.3). Furthermore,  in the slave-boson  formulation the natural
 mean-field decouplings usually have simpler structure at finite doping than
those in the so-called slave-fermion, Schwinger-boson formalism.\cite{aa}

Various RVB-type  mean-field states have been obtained in the slave-boson
formulation.\cite{lee2} The mean-field decoupling of the Hamiltonians (2.7) and
(2.8) may be realized by introducing the following mean-fields
\begin{equation}
 \chi _{ij} ^{\sigma } =<f_{i \sigma }^+f_{j \sigma }>; \end{equation}
\begin{equation}
H _{ij} =<h_{i}^+ h_j>. \end{equation}
By neglecting  higher order fluctuations around $\chi _{ij}^{\sigma}$
and $H_{ij}$, the mean-field versions of $H_{t-J}$ can be obtained in the form
of $H_{t-J}^{MF}=H_h^{MF}+H_s^{MF}$, where
\begin{equation}
H_h ^{MF}=-t\sum _{<ij> \sigma } \chi _{ji}^{\sigma} h_i^+ h_j +H. c.;
\end{equation}
\begin{equation}
H_s^{MF}=-\frac J 2 \sum _{<ij> \sigma \sigma'}
(\chi _{ji}^{\sigma '}+\frac t J H_{ji} ) f_{i\sigma }^+ f_{j\sigma }+ H.c.
\end{equation}
The constraint (2.6) should be simultaneously relaxed at this level to a
mean-field one $<h_i^+h_i>+\sum_{\sigma }<f_{i\sigma}^+f_{i\sigma}>=1$.

As shown by Eqs.(2.11) and (2.12),  both $H_h^{MF}$ and $H_s^{MF}$ have
similar hopping forms decided basically by one mean-field $\chi_{ij}^{\sigma}$
(usually $H_{ij} \propto  \chi _{ij}^{\sigma} $ at the saddle-point). Since
$\chi _{ij}^{\sigma}$ is responsible for both the hopping
and antiferromagnetic superexchange strengths in Eqs.(2.11) and (2.12), an
optimization of these two competing charge and spin processes may be relatively
easy within this framework.

A simplest mean-field state can be obtained by choosing
$\chi _{ij} ^{\sigma} = \chi_0 $ and $H_{ij} =H_0$, known as the uniform RVB
state.\cite{grilli}  Such a state recently has attracted intensive attention,
as one has been able to go beyond the mean-field (or saddle-point)
approximation by including
the phase fluctuations of $\chi _{ij}^{\sigma}  $ and $H_{ij}$ in terms of the
gauge-field description.\cite{larkin,lee,ioffe}
A lesson learned from the gauge theory approach  is that a singular
transverse gauge fluctuation could become very important, which implies that
spinon and holon are usually confined\cite{fradkin} by gauge field, in opposing
to the artificial decomposition of electron operator in  scheme (2.5).  The
non-conventional transport phenomena have been discussed\cite{lee,ioffe} in
this framework which are both theoretically and experimentally interesting.

The uniform RVB state is presumably energetically favorable at a
sufficiently large doping. On the other hand, when it is close to half-filling
 and the
doping concentration $\delta$ is low, the uniform RVB state may not be an
appropriate
saddle-point state because the antiferromagnetic energy could be
underestimated. For example, at half-filling limit with $\delta
\rightarrow 0$,
the variational superexchange energy of the uniform RVB state, which can be
computed by using the variational Monte Carlo ( VMC )  method with
the no-double-occupancy condition (2.6) being exactly implemented, is
found\cite{liang}
to be $-0.53J$ (per site), about $20\%$ higher than the best ground state
energy $-0.6692J$ [Ref.\onlinecite{trivedi}].
Furthermore, no AF long-range order would appear in
this limit as it should be present in the true ground state of $H_{t-J}$.

A substantial improvement of the variational ground-state energy at
half-filling can be realized by introducing a phase to $\chi _{ij} ^{\sigma} $
in Eq.(2.9), i.e.,
\begin{equation}
\chi _{ij}^{\sigma}=\chi _0 e^{i\theta_{ij}} .\end{equation}
(Note that $H_{ij}=0$ at $\delta =0$.) Here the phase $\theta _{ij} $ can not
be simply reduced  to a difference like $\theta_i-\theta_j$, or in other words,
a summation of $\theta _{ij}$ along the closed bonds of a plaquette,
\begin{equation}
 \Phi_{\Box} ={\sum}_{\Box } \theta_{ij}     \end{equation}
is none zero. $\Phi_ {\Box} $ may be regarded as some fictitious flux
threading through the
plaquette. At half-filling, the variational energy is optimized at $\Phi _{\Box
}= \pi$, known as the $\pi $-flux phase,\cite{affleck} whose VMC  value of the
exchange energy is $-0.623 J$ [Ref.\onlinecite{liang}], only
$5\% $ higher than the exact ground-state energy. However, the absence of a
long-range AF order in the $\pi $-flux phase implies that the
long-wavelength, low-energy AF correlations are still underestimated
which should
predominantly account for the $5\%$
energy missing at this saddle-point state. On the other hand, the high-energy
properties of the Heisenberg model, like the temperature dependence of the
specific heat may  be well explained in terms of the $\pi $-flux phase,
as discussed in Ref.\onlinecite{wang} (In  Ref.\onlinecite{wang}, the
$\pi$-flux phase is obtained from a different formulation, involving a 2D
Jordan-Wigner transformation\cite{jw2d} of the spin-1 operators, instead of
the spin-$\frac1 2$ operators used in the present framework.)

Thus, by comparing with the uniform RVB state at half-filling, the
short-range AF correlation in the $\pi $-flux phase has been  substantially
improved as a result that the transverse gauge fluctuation in the former is
condensed
as the static uniform flux in the latter case. Nevertheless, lack of long-range
AF
order  in
the latter means that the long-range AF correlation is still not appropriately
accounted here. In the following, we will consider a new type of saddle-point
states which could retain the high-energy characteristics of the $\pi- $ flux
phase, while properly include the long-wavelength AF correlation in low-energy
regime.

\subsection{Flux binding: new saddle-point}

In the $\pi$-flux phase discussed above, the mean-field Hamiltonian
describes a
non-interacting fermion gas under some uniform fictitious magnetic
field. However, according to the exact diagonalizations on small
lattice,\cite{canright} such a
type of states is generally not energetically favorable as compared to a state
in which the uniformly-distributed lattice flux is quantized into
infinitesimal-size flux tubes, and each of them is bound to an individual
particle. The former may be regarded as a ``mean-field'' version of the latter.
In general, the latter is known as an anyon system\cite{laughlin} as the new
composites of particles and flux-tubes obey different statistics, depending on
the flux strength of each flux-tube, as compared to
the statistics of the  underline particles.

This numerical result is very suggestive here, as  in the slave-boson
formalism
there  exists a gauge (phase) degree of freedom which  can allow the above
flux binding procedure to happen if it is indeed energetically favorable. This
gauge freedom is manifested in the  decomposition (2.5), where one may always
associate phases $e^{i\theta}$ and $e^{-i\theta}$ to $h_i^+$ and $f_{i\sigma}$,
respectively, without changing $c_{i\sigma}$. Such a freedom reflects the
arbitrariness  of  the decomposition (2.5). A gauge field will then play a role
to confine any non-physical spinon and holon together as an electron. Only when
one finds a correct spin-charge decomposition,  the gauge
fluctuation can be expected to get suppressed. Our strategy in the following is
to optimize the flux phase by exploiting such a gauge freedom. Main  procedure
is to regard the uniform fictitious flux as a ``mean-field'' version of the
system where fluxes are quantized and bound to the particles. This latter
system may be generally called as a flux-binding state. Given the
above-discussed uniform
$\pi$-flux, there can be two ways to construct the
corresponding flux-binding states at half-filling, as will be outlined below.

{\em Scheme one} \hspace{0.3in} All of $f_{i
\sigma } $, no matter their spins, are bound to the same type of flux quanta.
When these flux quanta are uniformly smeared out in space, one should recover
the  uniform flux in  $\pi $-flux phase.
At half-filling, the total number of spinons is one per plaquette on average.
Thus  each flux tube has to be quantized at $ \pi $, such that
\begin{equation}
{\sum}_{\Box } \theta _{ij} =\pi \sum _{l\in \Box}\sum_{\sigma  }
n_{l\sigma } ^f  \end{equation}
with $n_{l\sigma }^f =f_{l \sigma } ^+ f_{ l \sigma }$. On average
$<\sum _{\Box } \theta _{ij}>=\pi $. In this scheme, since each flux-tube is
quantized at $\pi $, spinons are
effectively turned into semions\cite{laughlin} (an exchange of two semions
gives rise to
a phase $\pm i$). {\em Scheme one} has been already discussed in detail in
Ref.\onlinecite{fb1}.  Away from half-filling, the mean-field version of this
flux-binding state
corresponds to the so-called commensurate flux-phase (CFP)
[Ref.\onlinecite{cfp}]. And
the gauge theories\cite{wiegmann,hasegawa,rod,zou2} based on CFP
has a close connection with such a flux-binding state.
In both cases, superconducting  condensation is found in the ground state, and
is related to the semionic condensation.\cite{laughlin} It has been shown in
the flux-binding state\cite{fb1} that there is no explicit time-reversal and
parity symmetry broken due to the cancellation between charge and spin degrees
of freedom. The most interesting features  show up in the normal
state,\cite{fb2} which can well explain  the anomalous in-plane
transport properties in cuprates, including  resistivity, the Hall effect, and
thermopower. But such a state still has trouble to recover the AF ordering at
half-filling.

{\em Scheme two}\hspace{0.3in} In this scheme, the spinons with
opposite spins may not see the flux-tubes carried by each other. Thus the flux
quantum of each flux-tube has to be doubled in order to recover the $\pi$-flux
phase in the mean-field limit. In this case, the phase $\theta _{ij}$ as
seen by the spinons with spin index $\sigma$ has to  satisfy the following
condition
\begin{equation}
{\sum}_{\Box}\theta_{ij}^{\sigma }=2 \pi \sum_{l\in\Box}n_{l \sigma}^f,
\end{equation}
such that $<\sum _{\Box } \theta _{ij} ^{\sigma } > =\pi $ at half-filling.
Equation (2.16) means that each spinon is bound to a $2\pi $ flux-tube, which
would effectively change the spinon statistics from fermion into (hard-core)
boson, for spinons with the same spin index $\sigma$.
In this second flux-binding scheme,\cite{fb3} the experimental features of the
transport properties found in {\em Scheme one} will be essentially
retained, while interesting magnetic properties can be also obtained here.
As an example, the AF long-range order can be recovered at half-filling.
Furthermore, the correct 1D behaviors is to be reproduced naturally within
this framework. All of these suggest that {\em Scheme two} may be a better
approximation than  {\em Scheme one} for the {\it same} physical state.
Therefore, we
shall fully focus on the saddle-point of {\em Scheme two} later, and explore
its various properties in the remainder of the paper.

\subsubsection{Mathematical scheme: Half-filling}

At half-filling, $H_t $ has no  contribution. One may rewrite
$H_J$ in Eq.(2.8) in a form
\begin{equation}
H_J  =-\frac J 2 \sum _{<ij> \sigma \beta } \left (e^{ i A_{ij}^{\sigma
}}f_{i\sigma } ^+ f_{j \sigma }\right)\left (e^{ i A_{ji}^{\beta}}
f_{j\beta} ^+ f_{i \beta}\right) ,    \end{equation}
where  a phase $A_{ij}^{\sigma } $, which will play a role similar to
$\theta_{ij}^{\sigma}$ in Eq.(2.16),  is introduced:
\begin{equation}
 A_{ij} ^{\sigma } =\sigma \sum _{l \neq i,j} [\theta_i(l)-\theta _{j} (l) ]
\left(n_{l \sigma } ^f - \delta _{\sigma, \uparrow } \right) , \end{equation}
which depends on the positions of spinons nonlocally. This nonlocality is
introduced through the multiple-valued phase $\theta _i (l) $:
\begin{equation}
 \theta _i (l) =Im \ln (z_i -z_l ), \end{equation}
with $z_i \equiv x_i+iy_i $ in 2D. A vorticity $\pm 2\pi $ can be obtained in
$\theta_i (l )$ if the coordinate $z_i $ has been continuously changed around
$z_l$ once. In 1D case, $\theta _i (l)$ in Eq.(2.19) will reduce to
\begin{equation}
\theta _i (l) =\pm \pi \theta (l-i),  \end{equation}
where $\theta (x) $ on the right-hand-side is the step function. In obtaining
Eq.(2.17), the constraint condition (2.6) has been used, i.e.,
\begin{equation}
 \sum _{\sigma }n_{i\sigma } ^f =1   ,        \end{equation}
at $\delta =0.$

The identical transformation in Eq.(2.17) has no real physical meaning until
the mean-field decoupling scheme is introduced. One may define a
hopping operator
\begin{equation}
\hat{\chi}_{ij}^{\sigma } =e^{ i A_{ij}^{\sigma  }}
f_{i\sigma } ^+ f_{j \sigma}  .\end{equation}
Then under the mean-field $\chi_0 =<\hat{\chi}_{ij}^{\sigma }>$, a
saddle-point Hamiltonian of Eq.(2.17) is found as
\begin{equation}
H_s=-J\chi _0 \sum _{<ij> \sigma }e^{i A_{ij}^{\sigma}}f_{i\sigma}^+
 f_{j\sigma}+H.c . \end{equation}
Now the nonlocal field $A_{ij} ^{\sigma }$ appears as a phase in the hopping
matrix, which is equivalent to the role of $\theta _{ij}$ in Sec.{\em A}.

{\em Two-dimensional case} \hspace{0.3in}In terms of Eqs.(2.18) and (2.19), a
summation of the phase $A_{ij} ^{\sigma } $ along a plaquette gives
\begin{equation} {\sum}_{\Box } A_{ij}^{\sigma }=2 \pi \sigma \sum _{ l \in
\Box } n_{l \sigma
}^f -2\pi \delta _{\sigma, \uparrow} .    \end{equation}
Besides a ($-2\pi $) lattice flux for $\sigma =\uparrow $, which has no real
physical meaning, the right-hand-side of Eq.(2.24) implies that each spinon
of
spin $\sigma $ carries a flux-tube quantized at $2 \pi $, with a sign $\sigma
=\pm 1$. Since $<n_{l\sigma } ^f> =1/2$,  one has
$ \sum_{\Box } <A_{ij}^{\sigma }>=-\pi$. That is, the ``mean-field '' version
of the saddle-point Hamiltonian (2.23) just corresponds to a $\pi $-flux phase
Hamiltonian as required (note that $-\pi$ and $\pi$ fluxes per plaquette are
equivalent).

As pointed out before, the flux binding procedure discussed here is
related to
a statistics-transmutation which is easy to see
after introducing a new operator $\bar{b}_{i \sigma }$
\begin{equation}
\bar{b}_{i \sigma }=f_{i\sigma } e^{-i\sigma \sum _{l\neq i }\theta_i (l)(n_{l
\sigma }^f-\delta _{\sigma \uparrow})} .\end{equation}
Then $H_s$ is reduced to
\begin{equation}
 H_s =-J \chi _0 \sum _{<ij > \sigma } (-\sigma )
 \bar{b}_{i \sigma }^+ \bar{b}_{j \sigma } +H.c., \end{equation}
One can easily check that  $\bar{b}_{i \sigma}$ satisfies the hard-core boson
commutation relations (for the same index $\sigma$ ): $[\bar{b}_{i \sigma }$,
$\bar{b}_{i \sigma }^+]=0$ ($i\neq j$), etc. Of course, for opposite spins  one
still finds anti-commutation relations.

Equation (2.25) resembles the Jordan-Wigner transformation in both 1D
and 2D cases\cite{jw,jw2d} which changes the statistics of a fermion into that
 of a hard-core
boson. Thus the presence of a uniform lattice flux in the $\pi$-flux
phase is a
precursor for a fermionic spinon to become a
boson. Such a ``bosonization'' tendency for spinons in flux phase has been
already noted before.\cite{hald2} Bosonic
representation has been known generally to be superior in the description of
spin AF correlations.

Equation (2.26) can be further written in the form of a standard hopping
Hamiltonian
\begin{equation}
H_s=-J\chi _0 \sum _{<ij > \sigma }  b_{i \sigma }^+b_{j
\sigma } + H.c. \end{equation}
by redefining
\begin{equation}
 b_{i \sigma } =(-\sigma )^i \ \ \bar{b}_{i \sigma },  \end{equation}
in which $\sigma =\pm 1$, and $(-1 )^i=+1$ for the even-sublattice  and
$-1$ for
the odd-sublattice of a bipartite lattice.  The spin operator like
$S_i^+=f_{i \uparrow}^+ f_{i \downarrow} $ can then expressed in terms $b_{i
\sigma}$ as follows
\begin{equation}
 S_i ^+ =(-1 )^ib_{i\uparrow }^+ b_{i\downarrow} .  \end{equation}
One also has $S_i ^-=(S_i ^+ )^+$, and $S_i ^z=\frac 1 2 \sum_{\sigma } \sigma
b_{i\sigma }^+b_{i \sigma }$.

The Bose-condensation of $b_{i \sigma } $ at zero-temperature as determined by
Eq.(2.27) means
\begin{equation}
< {S}_i ^+ > =(-1)^i <b_{i\uparrow}^+>   <b_{i\downarrow }> \neq 0.
\end{equation}
In other words, an antiferromagnetic long-range order in the x-y plane
can be indeed obtained in the present flux-binding scheme. The direction of
the  magnetization in x-y plane will be determined by the relative phase
between the up and down spinons in Eq.(2.30), and could be arbitrary. The
evaluation of the magnetization value
in Eq.(2.30) needs a detailed knowledge of the hard-core boson
behavior governed by $H_s $. Since we do not know the exact solution of the
hard-core boson Hamiltonian
(2.27), we can not directly estimate the variational ground state energy  of
the Heisenberg Hamiltonian $H_J$ in the present saddle-point state.
Nevertheless, the ground state energy for a hard-core boson system like
Eq.(2.27) is
generally lower
as compared to the uniform $\pi $-flux system in the fermion representation,
which has been discussed by Gros {\em et al.} in Ref.\onlinecite{gros}.
As pointed out by these authors, the ground state energy of a
hard-core boson system described by $H_s$ could be  as lower as
$11\% $  than that  in
the $\pi$ -flux-phase. Of course, one still needs to check the variational
energy of the {\em original} Hamiltonian $H_J$ in the present saddle-point
state. By
using a trial bosonic wavefunction for $H_s $ given in Ref.\onlinecite{gros},
which is
obtained by taking the absolute value of the known ferminic $\pi $-flux-phase
wave function and gives $7.4\%$ higher variational energy than the best
estimate for $H_s $,
we found analytically the variational energy of $H_J$ is just identical to that
in the $\pi $-flux phase. The latter is already known from the VMC calculation,
i.e., $5\%$ higher than the exact value of the Heisenberg Hamiltonian.
Since the
above-mentioned bosonic trial wavefunction does not have a long-range order
(Bose-condensation), one may attribute the $5\%$ higher ground state energy to
it. A better bosonic trial wavefunction of $H_s$ should further improve the
variational ground state energy of $H_J$.

We note that beyond
the present mean-field approximation, a gauge fluctuation should be considered.
As will be discussed in Sec. {\em C}, such a gauge fluctuation
is not gapped at half-filling and thus could become very important. In fact,
spin-$1/2$ spinons will presumably be confined by the gauge field to form
spin-$1$ excitation (spin wave) at half-filling. This bosonic description is
different from the Schwinger boson representation. The latter has been
found\cite{aa} to well describe the low-lying spin excitation in 2D
half-filling,  even at a mean-field level.
In contrast, the present scheme will become more powerful in spinon
deconfined cases. For instance, it will be shown later that the gauge
fluctuation is to be suppressed at finite doping, and with the presence of
spin-charge separation there, the present bosonic representation can become
quite convenient to include doping effect. An another spinon
deconfinemnet case
is in 1D, which will be discussed below.

{\em One dimensional case}\hspace{0.3in} As a further check of the present
saddle-point state, we now turn to the one dimensional case where
the asymptotic spin-spin correlation of the Heisenberg chain has been
known for
many years.\cite{luther}

Under the saddle-point  Hamiltonian (2.27), one can write
\begin{equation}
<S_i^+(t)S_j^- (0)>=(-1)^{i-j} <b_{i\uparrow}^+(t) b_{j\uparrow} (0)>
<b_{i\downarrow}(t) b_{j\downarrow}^+ (0)>  . \end{equation}
In order to evaluate the averages on the right-hand-side, the behavior of the
hard-core boson described by $b_i (t) $ has to be determined first. A trick
which has been often used in 1D is to transform the bosonic operator $b_i$ into
fermionic operator again. By using the expression (2.20) for $\theta _i (l) $
in 1D, $b_{i \sigma } $ is expressed in terms of $f_{i \sigma } $ in Eqs.(2.25)
and (2.28) as follows
\begin{equation}
 b_{i \sigma } =f_{i \sigma } e^{\mp i \sigma \pi \sum _{l > i } n_{l \sigma }
^f }              .
\end{equation}
One also finds that $A_{ij}^{\sigma } =0 $ in Eq.(2.23), and thus in the
fermionic representation
$H_s $ simply describes a free lattice fermion gas. Then the correlation
function like $<b_{i \uparrow }^+ (t) b_{j \uparrow } (0)  >$ in
Eq.(2.31) can
be straightforwardly calculated in long-time and -distance as
follows\cite{affleck2}
\begin{eqnarray}
 <b_{i \uparrow }^+ (t) b_{j \uparrow } (0)  >=&&<  f_{i\uparrow
}^+ (t)
e^{\pm i \pi \sum _{l > i} n_{ l \uparrow }^f (t)}
e^{\mp i \pi \sum _{l > j} n_{ l \uparrow }^f (0) } f_{j \uparrow }
(0)>\nonumber  \\
&&\propto \frac 1 { [(x_i-x_j)^2-v_f^2t^2]^{1/4}  }  ,
\end{eqnarray}
with $v_f= J\chi_0 a       $. The average $ <b_{i\downarrow } (t) b_{j
\downarrow }
^+ (0) > $ shows
the same asymptotic form. Finally by using $(-1)^{i-j} =cos \frac \pi a
(x_i -x_j ) $ and the rotational invariance, one obtains
\begin{equation}
<{\bf S}_i(t)\cdot {\bf S}_j(0)> \propto \frac {cos \frac {\pi } a (x_i-x_j)}
{ [(x_i-x_j)^2-v_f^2t^2]^{1/2}  }         \end{equation}
which describes the correct asymptotic spin-spin correlation in 1D and was
first derived by Luther and Peschel from XXZ model.\cite{luther}

Thus the present flux-binding saddle-point state has produced a correct
result in 1D,  even though it is originally constructed in 2D. This is in
contrast to the
Schwinger-boson approach,\cite{aa} which works rather successfully for the 2D
Heisenberg model, but
could not predict the correct behavior for the 1D spin 1/2 case. This may be
attributed to the fact that spin-$1/2$ excitations exist in 1D
Heisenberg chain,
but not in 2D. The present bosonic representation properly describes these
spin-$1/2$ spinons.

We note that in the present approach the spin-spin correlation in
$z$-axis can not be evaluated at the same level of approximation as in the
$x-y$ plane. The reason is that
$<S_i^z\cdot S_j^z>\sim \frac{1}{4}\sum_{\sigma}<n_{i\sigma}^bn_{j\sigma}^b>$
basically involves the
density-density correlation of spinons with the same spin index, while
$<S_i^+\cdot
S_j^->$  is related to the single-particle  propagators
of spinons under the saddle-point (2.27). The former is generally much harder
to evaluate because of the hard-core condition of $b_{i\sigma}$.
The 1D correlation  $<S_i^z\cdot S_j^z>\sim (-1)^{i-j}/|x_i-x_j|$ could be
obtained with the hard-core condition being accurately treated through the
Jordan-Wigner transformation.

One final point we like to stress here is that $H_s$ in Eq.(2.23) is
formally the same as that in the uniform RVB saddle-point, since $A_{ij}
^{\sigma } =0$ in 1D. In other words, the flux-binding, or the
statistics-transmutation, has no effect on the Hamiltonian in 1D as  is well
known. What is different between these two saddle-point states is the boundary
condition for the fermion $f_{i \sigma }$: in the present saddle-point, $f_{i
\sigma }$ appearing in Eq.(2.32) will become multiple-valued at the
boundary once the 1D chain is closed as a loop.   As discussed by
Shastry,\cite{shastry2}
such a boundary condition decides why a hard-core boson system is energetically
more favorable
than a fermionic gas.

\subsubsection{Flux-binding:  finite doping}

For a small doping concentration, one expects the
antiferromagnetic correlation still to remain dominant, even though it may  be
strongly modified by doped holes. It is very interesting  to see the evolution
of the flux-binding saddle-point state in this finite doping case.

We may follow the procedure shown in Sec. {\em 1}, starting with the identical
transformation of $H_J$ in Eq.(2.17). At finite doping, the form of
$A_{ij}^{\sigma} $ in  Eq.(2.18) will be slightly modified because the
constraint  Eq.(2.21) now is replaced by the full equation (2.6). It is easy to
show that $A_{ij}^{\sigma} $ is given as follows:
\begin{equation}
A_{ij}^{\sigma }=\sigma  \sum_{l \neq i,j} [\theta _i (l)- \theta _j (l)]
\left( n_{l \sigma }^f -\delta _{\sigma,\uparrow } +\frac 1 2 n_{l} ^h\right)
,\end{equation}
which keeps the identical transformation in Eq.(2.17) unchanged. In Eq.(2.35),
an additional factor $ n_l^h /2$ is introduced, with $n_l ^h=h_l^+ h_l $ as
the holon number operator.

At finite doping, the hopping  term $H_t $ in the $H_{t-J}$ will also
contribute, and it may be rewritten in a form
\begin{equation}
H_t=-t \sum_{<ij>\sigma} e^{i A_{ij}^f } h_i^+ h_j \hat \chi_{ji}^{\sigma },
\end{equation}
where $\hat{\chi}_{ji}^{\sigma } $ is defined by Eq.(2.22) with $A_{ji} ^{
\sigma } $ given in Eq.(2.35). $A_{ij}^f $ in Eq.(2.36) is a re-arranged form
of $A_{ij}^{\sigma } $ after  using
the constraint (2.6):
\begin{equation}
A_{ij}^f =\frac 1 2 \sum_{l \neq {i,j} } [\theta _i (l)- \theta _j (l)]
\left(\sum _{\sigma } \sigma n_{l \sigma} ^f -1\right).  \end{equation}
Note that $A_{ij} ^f$ in Eq.(2.37) has  no explicit $\sigma $-dependence.

Now we may take the mean-field decoupling of $H_J $ in Eq.(2.17) and $H_t $ in
Eq.(2.36), by introducing the mean-fields $\chi_0 =<\hat \chi _{ij}^{\sigma } >
$ and  $H_0 =<\hat H_{ij}> $ (here $\hat H_{ij} \equiv e^{i A_{ij}^f } h_i^+
h_j $). At
the same time, the constraint (2.6) has to be relaxed up to a global
level, as outlined before. Then a saddle-point Hamiltonian at
finite doping is found to be
\begin{equation}
H_{t-J} ^{MF} =H_s+H_h ,\end{equation}
in which the spinon degree of freedom has the same form
as in the half-filling case:
\begin{equation}
H_s =-J_s \sum _{<ij>, \sigma } e^{i A_{ij}^{\sigma  } }f_{i \sigma }^+
f_{j \sigma } +H.c.   ,\end{equation}
with $J_s =J\chi_0+t H_0$. The holon degree of freedom is described by:
\begin{equation}
H_h =-t_h \sum _{<ij>} e^{i A_{ij}^f } h_{i }^+ h_{j } +H.c.  ,\end{equation}
with $t_h=2t \chi _0$. $t_h$ and $J_s$ have to be determined self-consistently.
 Since the averages $\chi _0$ and $H_0$ actually
do not depend on $t_h$ and $J_s$, the self-consistency should be always
satisfied here,
even though an actual determination of the values for $t_h$ and $J_s$ is
nontrivial.  Generally one can estimate  $t_h\sim t$ and $J_s\sim J$ at
small doping. In the bosonic representation
of Eqs.(2.25) and (2.28), $H_s $ can be further expressed as
\begin{equation}
H_s =-J_s  \sum _{<ij> \sigma } e^{i \sigma  A_{ij}^h } b_{i \sigma }^+
b_{j \sigma } +H.c.     ,   \end{equation}
where the topological phase $A_{ij}^h $ is defined by
\begin{equation}
A_{ij}^{h }=\frac 1 2   \sum_{l \neq i,j}\left[\theta _i (l)- \theta _j (l)
\right]\hspace{0.01in} n_{l }^h    .\end{equation}

In one dimension, one will find $A_{ij}^h=A_{ij}^f =0 $ in Eqs. (2.40) and
(2.41) according to Eq.(2.20).
The holon (described by $h_i $) and spin (described by $b_{i \sigma } $)
degrees of freedom are thus decoupled at this saddle-point state. In two
dimensions, however, the holon and spinon are coupled with each other through
the nonlocal fields $A_{ij}^f $ and $A_{ij}^h $ in $H_h $ and
$H_s$. The topological gauge field $A_{ij}^h$ can be interpreted as describing
fictitious $\pi$ flux quanta
attached to holons, but only seen by spinons. This becomes clear if one
considers an arbitrary loop $C$, and counts the flux enclosed, which is
given by
\begin{equation}
{\sum}_C A_{ij}^h=\pi \sum_{l\in C} n_l^h .   \end{equation}
$A_{ij}^f $ in Eq.(2.37) may be rewritten as $A_{ij}^f = A_{ij}^s-\phi_{ij}^0
$, where
\begin{equation}
A_{ij}^s =\frac 1 2 \sum_{l \neq {i,j} } [\theta _i (l)- \theta _j (l)]
\left(\sum _{\sigma } \sigma n_{l \sigma}^b\right),  \end{equation}
and $\phi_{ij}^0$ is defined by
\begin{equation}
\phi_{ij}^0 =\frac 1 2   \sum_{l \neq i,j}\left[\theta _i (l)- \theta _j (l)
\right]  , \end{equation}
which describes  a lattice $\pi$-flux with $\sum_{\Box} \phi_{ij}^0=\pi$.
The topological gauge filed $A_{ij}^s$
describes  fictitious $\pi $ flux-tubes carried by spinons which are seen by
holons. In Secs. III and IV, we will investigate how
the couplings induced by nonlocal fields $A_{ij}^h $ and $A_{ij}^s $ will
lead to
highly nontrivial spin and charge properties in 2D case.

To end this section, we would like to give a different perspective about the
construction of the flux-binding state. Recall that one has a gauge
degree of freedom in a slave-boson decomposition: $c_{i\sigma}=h_i^+b_{i\sigma}
=\left(h_i^+e^{i\theta_h}\right)\left(f_{i\sigma}e^{i\theta_f}\right)$, where
$\theta_h$ and $ \theta_f$ can be {\em any} phase satisfying
$\theta_h+\theta_f=0$.
Particularly, if $\theta_h$ and $\theta_f$ are chosen as the
topological phases
in the Jordan-Wigner transformation, statistics-transmutation can happen
so that
one may end up with a slave-fermion, or more generally, slave-anyon
decomposition.\cite{fb1} This is not
surprising because all of these formalisms are mathematically equivalent. Much
complicate decomposition can be also constructed. Distinctive physics is
involved here
only when one makes the mean-field decoupling. In this way, different
decomposition
means different saddle-point state. The usual slave-boson and slave-fermion
formalisms
are the simplest ones, but not necessarily the appropriate ones for the
correct saddle-point. The present saddle-point state corresponds to
$\theta_h=1/2\sum_{l\neq i}\theta_i(l)[\sigma(1-n^h_l)+\sum_{\alpha}\alpha
n_{l\alpha}^f]$ and $\theta_f=-i\sigma \sum_{l\neq i} \theta_i(l)
n_{l\sigma}^f$.
It is easy to check that  $\theta_h+\theta_f=0$    in terms of the
constraint (2.6).
Correspondingly, $c_{i\sigma}$ may be rewritten as in Eq.(1.6) where
$b_{i\sigma}$
is defined  by Eqs.(2.25) and (2.28). In this new decomposition, an
electron is
composed of two {\em bosonic} holon and spinon, together with some nonlocal
fields.  The spin operator like
$S^+_i=f_{i\uparrow}^+f_{i\downarrow}$ can also be expressed as
\begin{equation}
S^+_i= (-1)^i b^+_{i\uparrow}b_{i\downarrow} e^{i\sum_{l\neq i}\theta_i(l)
n_l^h}.
\end{equation}
Its physical meaning will be explored in Sec. III. In the following, we will
go beyond the mean-field approximation and show that a real spin-charge
separation can be indeed realized with the  decomposition (1.6).

\subsection{Gauge-theory description: Spin-charge separation}

It is now a well-recognized fact\cite{lee,ioffe} that any saddle-point-state
properties of
the $t-J$ model could be substantially modified by the low-lying gauge
fluctuation around the saddle-point. So it is very important to check the
gauge fluctuations around the present saddle-point.
Here  gauge fluctuations are to be related to phase fluctuations of
the mean-fields $<\hat \chi _{ij}> $ and $ < \hat H_{ij}> $.  The main idea
behind the construction of  flux-binding
saddle-point states  is to incorporate the most singular transverse
gauge fluctuations into the saddle-point via flux binding so that additional
phase fluctuations become insignificant. These singular gauge fields
are represented by the topological gauge fields $A_{ij}^h$ and $A_{ij}^s$ in
the saddle-point Hamiltonians $H_s$ and $H_h$. In the following, we will
show that the transverse gauge fluctuation around this saddle-point is indeed
suppressed  at finite doping, and therefore  we have a real spin-charge
separation whose low-energy physics will be determined in terms of the
effective Hamiltonians $H_s$ and $H_h$, defined by Eqs.(1.1)-(1.5) in
the Introduction.

An  effective
Lagrangian, with the constraint (2.6)  enforced and  the (phase) gauge
fluctuation around the present flux-binding saddle-point state included, can be
written
down as\cite{lee,fb1}
\begin{equation}
{\cal L}_{eff}=\sum_i\left(\sum_{\sigma}b_{i\sigma}^+(\partial_{\tau}-\mu)b_{i
\sigma}+h_i^+(\partial_{\tau}-\mu)h_i\right) + \tilde{H}_{eff}, \end{equation}
where
\begin{eqnarray}
\tilde{H}_{eff}=&&-J_s\sum_{<ij>\sigma} e^{i(a_{ij}+ \sigma
A_{ij}^h)}b_{i\sigma}^+b_{j
\sigma} + c.c. \nonumber
\\         && -t_h \sum_{<ij>} e^{i(a_{ij}+A_{ij}^s
-\phi_{ij}^0)}h_i^+h_j + c.c.
 \end{eqnarray}
Here $a_{ij}$ in Eq.(2.47) describe
an internal gauge field. Originally the no-double-occupancy
constraint Eq.(2.6) [with $f_{i\sigma}$ changed to $b_{i\sigma}$ by
transformations Eqs. (2.25) and (2.28)] is implemented through a Lagrangian
multiplier field, whose
fluctuating part is then absorbed by the longitudinal part of $a_{ij}$ in
Eq.(2.48) due to the gauge invariance,\cite{fb1} with $\mu$ left in Eq.(2.47)
enforcing
the constraint at a global level.

The existence of a gauge freedom in ${\cal L}_{eff}$  will guarantee the
following current constraint between spinon and holon:\cite{lee,fb1}
\begin{eqnarray}
{\bf J}_s=-{\bf J}_h    . \end{eqnarray}
It is noted that Eq.(2.49) in longitudinal channel is a simple
reflection of the
density constraint (2.6), but its transverse channel has nothing directly to
do with the no-double-occupancy constraint, and is related to the property of
the $t-J$ Hamiltonian. It is the transverse gauge fluctuation that becomes
singular in long-wavelength, low-energy regime in the uniform RVB
state.\cite{lee,ioffe}
This transverse field would  serve as a confine force in a usual
non-spin-charge-separation state. The current constraint (2.49) is also
connected to the Ioffe-Larkin combination rule\cite{larkin} of the response
to an external electromagnetic field
\begin{equation}
K_e= [ K_s^{-1}+ K_h^{-1} ]^{-1} , \end{equation}
where $K_s$ and $K_h$ are the response matrices of spinon and holon systems
under effective Hamiltonians $H_s$ and $H_h$ in Eqs.(1.2) and (1.3).

In gauge theory, the dynamics of the gauge field $a_{ij}$ is determined after
the spinon and holon degrees of freedom are integrated out.\cite{larkin,lee}
When the gauge
fluctuation is weak, one may use the Gaussian approximation, and the gauge
field propagator $D^a_{\mu\nu}=-<T_{\tau} a_{\mu} a_{\nu}>$ ($\mu$, $\nu = x$,
$y$) in imaginary-time can be determined by\cite{larkin,lee}
\begin{equation}
D^a=-[K_s+K_h]^{-1}  ,  \end{equation}
in imaginary-frequency space.  In the present system, holons are under some
fluctuating flux described by $A_{ij}^s$ [cf. Sec. IV], and its response
function will follow  a usual metallic  ${\bf q}$ and $\omega$
dependances:\cite{larkin}
$i\omega \sigma_h-\chi_h q^2$. The most interesting behavior will come from
the spinon part $K_s$ as discussed below.

In the spinon part of $\tilde{H}_{eff}$, there is a sign $\sigma=\pm 1$ in
front of the topological phase $A_{ij}^h$. This sign means that spinons with
$\uparrow$ and $\downarrow$ spins see the fictitious flux quanta carried by
holons in opposite directions. Since there is no such a sign in front of the
gauge field $a_{ij}$, a non-zero $a_{ij}$ will then polarize the spinon system
with regard to spin $\sigma$. This polarization is generally energetically
unfavorable, and we expect a suppression of  the gauge field $a_{ij}$ to
stabilize the system.

A mathematical demonstration is straightforward. As will be discussed in Sec.
III,  one may rewrite $A_{ij}^h$ as  $A_{ij}^h={\bar A}^h_{ij} + \delta
A_{ij}^h$, where ${\bar A}^h_{ij}$ is a ``mean-field'' one with the flux-quanta
uniformly smeared out in space. A uniform fictitious magnetic field $B_h=\pi
\delta/{\mbox a}^2$ will correspond
to such a  vector potential ${\bar A}^h_{ij}$. On the other hand, $\delta
A_{ij}^h$ will be
correlated with the holon density fluctuation. If one integrates out the spinon
degree of freedom (under the ``mean-field'' ${\bar A}^h_{ij}$) in ${\cal L}_{
eff}$, an effective
action for quadratic fluctuation of the gauge field is found by
\begin{equation}
S_s[a, \delta A^h]=\frac 1 2 \sum_{\sigma} \left(a +\sigma \delta A^h\right)
\Pi_{\sigma} \left(a +\sigma \delta A^h\right) ,   \end{equation}
where $\Pi_{\sigma}$ is the response matrix of spinon $\sigma$ under a
fictitious magnetic field $\sigma B^h$.  $K_s$ will be determined through
$S_s= \frac 1 2 a K_s a$ [Ref.\onlinecite{wiegmann}], and one needs to further
integrate
out $\delta A^h$
in Eq.(2.52). At first step, if one neglects $\delta A^h$'s own dynamics, by
directly integrating out $\delta A^h$ in Eq.(2.52),  the following expression
for $K_s$ can be obtained
\begin{equation}
K_s=4\left[\Pi^{-1}_{\uparrow}+ \Pi^{-1}_{\downarrow}\right]^{-1}   .
\end{equation}
The key thing here is that the up and down spinons see the opposite fictitious
magnetic fields, and the Hall conductance terms in $\Pi_{\uparrow}$ and
$\Pi_{\downarrow}$ will have opposite signs. As shown by
Wiegmann\cite{wiegmann} in a
generic
situation, due to the same amplitude but opposite signs of the Hall
conductances, $K_s$ in Eq.(2.53) must exhibit the ``Meissner effect'' in the
transverse channel: i.e.,  $K_s^{\perp}= 1/4\pi\lambda^2 \neq 0$ at $\omega=0$,
and ${\bf q}\rightarrow
0$ limit. Here we find $1/\lambda^2=\pi B_h J_s$. Such  a ``Meissner effect''
implies that the gauge field will get
suppressed in the spinon system. In the long-wavelength and  low-energy limit,
$K_s$ becomes
dominant in the transverse channel of $D^a$ as $K_h\sim 0$, and a non-zero
$K_s^{\perp}$ then leads to a gap $\propto \delta J$
in the transverse gauge fluctuation of Eq.(2.51) (Anderson-Higgs mechanism).
This gap
means  a spin-charge deconfinement as pointed out before. In contrast,
a finite $K_s^{-1}$ will play  a {\em negligible} role in
the electromagnetic response function $K_e$ in Eq.(2.50) at small
${\bf q}$ and $\omega$, and one finds
\begin{equation}
K_e\simeq K_h  ,
\end{equation}
in such a limit.
So the  holon degree of freedom solely determines the long-wavelength,
low-energy response to an
external electromagnetic field in this spin-charge separation system.
Obviously, all these features will disappear at $\delta\rightarrow 0$ as $B_h$
vanishes.

The above conclusion is still true if
one includes $\delta A^h$'s own dynamics by
adding a term $ \frac 1 2 \delta A^h {D^A}^{-1} \delta A^h$  in Eq.(2.52).
Here $D^{A}$ is
the free propagator for $\delta A^h$ which is related to holon density
fluctuation as
follows:
\begin{equation}
D^A_{\mu\nu}=\left(\delta_{\mu\nu}-\frac{q_{\mu}q_{\nu}}{q^2}\right) \left(
\frac{\pi^2}{q^2} D_{\rho}^h\right),  \end{equation}
with $D_{\rho}^h$ as the holon density-density correlation function. Then
Eq.(2.53) is modified as
\begin{eqnarray}
K_s=&& \left[ (\Pi_{\uparrow} + {D^A}^{-1})^{-1} +
\Pi_{\downarrow}^{-1}\right]^{-1}+
\left[ \Pi_{\uparrow}^{-1}+ (\Pi_{\downarrow} +
{D^A}^{-1})^{-1}\right]^{-1}\nonumber \\
&& +\left[ \Pi_{\uparrow}^{-1} + \Pi_{\downarrow}^{-1}+
\Pi_{\uparrow}^{-1}(D^A)^{-1}\Pi_{\downarrow}^{-1} \right]^{-1}+
\left[ \Pi_{\uparrow}^{-1} + \Pi_{\downarrow}^{-1}+
\Pi_{\downarrow}^{-1}(D^A)^{-1}\Pi_{\uparrow}^{-1} \right]^{-1}  .
\end{eqnarray}
Since $D^A$ does not have a transverse-longitudinal mixed term (the Hall term),
the above conclusion about $K_s$ can be easily found to remain unchanged as
long as $D^A$ approaches to a constant or vanishes slower than $q^2$ when
$q\rightarrow 0$. (Note
that the density-density correlation function\cite{tklee} of a hard-core boson
gas is very similar to a free-fermion gas, and in the latter case one has
$D_{\rho}^h\sim const.$ at $\omega=0$ and $q\rightarrow 0$.)

\section{SPIN DYNAMICS AT FINITE DOPING}

Spin dynamics in the insulating cuprates has been well understood
within the framework\cite{heisenberg} of the Heisenberg model.  A real
important
issue is
how the antiferromagnet is affected under doping. In this section, we
will  explore such a doping effect under the spin-charge separation
scheme, and
compare its unique features with those found in the cuprates.
We first  consider the 1D case, where the exact analytical results are
available for comparison.

\subsection{Spin-spin correlation in one dimension}

The present case corresponds to the large-$U$ limit of the Hubbard model, whose
exact solution\cite{lieb} has been obtained a long time ago. However, only
recently the
underlying physics and various correlation functions have been clarified,
following Anderson\cite{anderson,and1}and other
authors.\cite{ogata,frahm,schulz,weng1}

The spin-spin correlation function is one of important correlation
functions in 1D. According to the numerical calculation\cite{ogata} based
on the exact solution, a $2k_f$ oscillation is present in the spin-spin
correlation function.  One may naively  relate this incommensurate structure
with the existence of a large electron Fermi surface (points at $k_f$ and
$-k_f$).  Similar argument may also be applied to a 2D case. This observation
should be physically reasonable, but by itself is not sufficient to get the
correct spin dynamics. Much rich
effects will be involved here due to strong electron-electron correlations. In
fact, one has  a spin-charge separation. In this case, the
spinons are presumably responsible for the spin dynamics.  But at the same
time, holons as  solitons of doped holes bound with  spin domain-walls will
also
contribute to spin properties.  A combination of these decides a peculiar doped
spin-spin correlation, which is highly non-trivial from a Fermi-liquid point of
view. In the following, we will show how a correct spin-spin correlation can be
obtained without directly involving the electron Fermi-surface.

In the present scheme,  the spin flip operator $S_i ^+$ in Eq.(2.46) can be
expressed in 1D in terms of Eq.(2.20) as
\begin{equation} S_i ^+=(-1)^i b_{i \uparrow}^+  b_{i \downarrow}e^{\pm i \pi
\sum_{l >i}
n_l^h}. \end{equation}
Here a prominent feature is a Jordan-Wigner-type
nonlocal phase $\pi  \sum_{l >i} n_l^h      $ which involves holon number
operator. At first glimpse, one may feel strange for a holon number operator to
appear in a spin  expression. But this is not new, and   has already been
found in Ref.\onlinecite{weng1} by a path-integral approach, as a result of
spin-charge separation.  It actually describes the effect of spin domain-walls
carried by holons mentioned above.  This is
easy to see if one freezes the dynamics of spinon $b_{i\sigma}$
by letting it ``condensed'' in Eq.(3.1). Then one  would have  a
N{\'e}el order in spin $x-y$ plane when holons are absent. For each added holon
at $l$ site,
one finds  an extra sign $e^{\pm i\pi}=-1$ for every $S^+_i$ at $i<l$, which
means a flip for all those spins. In other words,  there is indeed a spin
domain-wall accompanying with each doped hole, and that is the physical reason
why the holon number operator enters into Eq.(3.1) nonlocally. We note that
the nonlocal field appearing in Eq.(3.1) can be also  related to
the phase-shift field that leads to  the Luttinger liquid behavior and an
electron Fermi-surface satisfying the Luttinger liquid theorem in the single
electron propagator, which is to  be discussed elsewhere.  In fact, a $2k_f$
oscillation
in the spin correlation
function will arise naturally due to this nonlocal field as shown below.

Due to   $A_{ij}^s= A_{ij}^h=0$ in 1D,  holon  and spinon  are decoupled in
the  Hamiltonians (1.2) and (1.3). The nonlocal phase in Eq.(3.1) will then
solely determine the doping effect. One may write
\begin{eqnarray} <S_i ^+(t)S_j ^-(0)>=&&(-1)^{i-j}< b_{i \uparrow}^+(t)
b_{j \uparrow}(0)> < b_{j \downarrow}(t)
b_{j \downarrow}^+(0)>\nonumber\\
&&\times < e^{\pm i\pi \sum_{l >i} n_l^h(t)}
e^{\mp i\pi \sum_{l >j} n_l^h(0)}> .
\end{eqnarray}
Since $b_{i \sigma } $ and $ h_i $ as described by Eqs.(1.2) and (1.3) are
hard-core bosons, each average on the right-hand-side of Eq.(3.2) can be
evaluated  straightforwardly in the asymptotic limit (see Sec. II {\em B}, and
Ref.\onlinecite{weng1}).
For example
\begin{equation}
 < e^{\pm i\pi \sum_{l >i} n_l^h(t)}
e^{\mp i\pi \sum_{l >j} n_l^h(0)}> \propto \frac {cos(\frac {\pi} {2a}\delta
x)}{(x^2-v_h^2t^2)^{1/4}}   ,   \end{equation}
with $x=x_i-x_j$, and $v_h= 2t_ha \sin (\pi\delta)     $.  The final result is
\begin{equation}
<S_i ^+(t)S_j ^-(0)>\propto   \frac {cos(2k_fx)}
{ (x^2-v_s^2t^2)^{1/2}(x^2-v_h^2t^2)^{1/4}} ,     \end{equation}
where $k_f=(1-\delta)\pi/2a$. The doping effect as
described by the phase-shift in Eq.(3.1) enters Eq.(3.4) by changing the
oscillation  from $2\pi/a$ to the incommensurate $2k_f$,  and at the same time
contributing an additional power
$(x^2-v_h^2t^2)^{-1/4}$. It is noted that $<S_i ^z(t)S_j ^z(0)>$ can not be
directly computed at the same level of
approximation as explained in Sec. II, but it should follows the same behavior
due to the rotational invariance.

Equation (3.4) recovers the correct spin behavior at the
strong-coupling fixed point of the model (i.e., the large-$U$ Hubbard model).
It suggests that the present state indeed catches the correct characteristics
of charge-spin separation in 1D, even though such a scheme  is originally
constructed for 2D.

\subsection{Doping effect in two dimensions}

In the one-dimensional case, spinon and holon are decoupled, and the
doping effect on spin dynamics is solely contributed by  the spin domain walls
carried by holons. In 2D, this nonlocal doping effect takes a different
form,
because spinon  and holon will no longer behave like  free solitons.
As noted before, the topological phase $A_{ij}^h $  in $H_s$ cannot
be gauged away in 2D, and  it will represent a new type of nonlocal influence
of doping on spinons.

\subsubsection{New length and energy scales introduced by doping}

As shown in Sec.II, the gauge field confining spinon and holon are suppressed
at finite doping.  But there are still  residual interactions between spinons
and holons, and
spinons can always feel the existence of holons nonlocally by seeing the
flux-quanta bound to the latter.

Let us first see  how this exotic interaction can change the topology of a
holon. In the spin flip operator $S_i^+$ (Eq.(2.46)), one  has a nonlocal phase
[$\sum_{l\neq i} \mbox {Im ln }(z_i-z_l) n_l^h)$]
involving holon number operator.  In resemblance to the domain-wall picture in
1D,  one may interpret it as describing
a spin vortex in $x-y$ plane with vorticity $2\pi$ bound to each holon. This
could be
seen if  one freezes the dynamics of  $b_{i\sigma}^+$ and treats it as a number
in Eq.(2.46). But $b_{i\sigma}^+$ here can no longer  be  regarded  as a
single-valued quantity because it is under the
topological phase   $A_{ij}^h$ in $H_s$.  A Berry-phase counting
shows that the phase in $b_{i\sigma}^+$ cancels out the effect of the
nonlocal phase-shift in Eq.(2.46) such that there is actually no
$2\pi$-vortex
topological texture formed around a holon. Nevertheless, if one goes  along a
line {\em across} such a  holon, one can  still find a domain-wall-like spin
singularity at holon site similar to that in 1D. Thus, a holon in 2D is no
longer associated with a spin topological object. Instead,
it may carry  a Shraiman-Siggia-type\cite{ss1} dipolar texture with vorticity
$=0$.  Such an object will not break the $T$- and $P$-symmetries. One expects a
strong dynamical renormalization to be involved in determining the profile of
each holon. However,  we are not  interested in a  single-doped-hole problem
here.
We shall focus on  a finite doping concentration in the following,  where a
useful mathematical description becomes available.

For simplicity, we are going to use  a continuum version of the spinon
Hamiltonian $H_s$ by taking the lattice constant $a\rightarrow 0$. This
continuum approximation may be justified at small doping and in the low-energy,
long-wavelength regime, where the  amplitude of
$A_{ij}^h$  is small and the spinons as bosons  mainly stay near the
bottom of energy band.
Such a continuum version for  $H_s $ in Eq.(1.2) can be easily written down as
[cf. Ref.\onlinecite{fb1}]
\begin{equation}
\tilde{ H}_s=\sum_{\sigma}\int d^2 {\bf r}\ \ b_{\sigma }^+ ({\bf r})
\frac {(-i\nabla -\sigma {\bf A}^h  )^2 } {2 m_s} b_{\sigma } ({\bf r}),
\end{equation}
with $m_s=(2J_s a^2)^{-1}$ and $ {\bf A}^h({\bf r})$ defined by
\begin{equation}
{\bf A}^h({\bf r})=\frac 1  2 \int d^2 {\bf r}' \frac {\hat{\bf z} \times (
{\bf r}- {\bf r}')}
{ | {\bf r}- {\bf r}' |^2}\rho _h ({\bf r}') . \end{equation}
Here $\rho _h ({\bf r}) $  is the holon density $\rho _h =h^+({\bf
 r})h({\bf r})$.

We note that Eqs.(3.5) and (3.6) would describe a semion problem\cite{laughlin}
if the holon density  $\rho_h ({\bf r}')$  were replaced by the spinon density.
The mathematical  similarity
suggests that one may borrow the method developed  in the  anyon
problem.\cite{laughlin,laughlin2,chen} The
idea is to rewrite $\rho _h =\bar { \rho _h}+(\rho_h -\bar { \rho_h})$ in
Eq.(3.6) by introducing an average holon density $\bar { \rho _h}=\delta
/{a^2} $.   Then ${\bf A}^h ({\bf r}) $ is rewritten as
\begin{equation} {\bf A}^h ({\bf r})={\bar {\bf A}}^h ({\bf r})+\delta {\bf
A}^h
({\bf r})
\ \ .  \end{equation}
Here ${\bar {\bf A}}^h (r)$ corresponds to $\bar {\rho }_h $, and in the
symmetric gauge it may be expressed by\cite{laughlin2,chen}
${\bar {\bf A}}^h (r)=\frac {B_h} 2 (\hat{{\bf z}}
\times {\bf r}) $ with $B_h=\pi \bar {\rho _h } $.  Physically, the vector
potential  $ \bar {\bf A}^h  $ describes a mean-field magnetic field $B_h $
obtained after the flux quanta bound to  holons are uniformly smeared out in
space. $\delta {\bf A}^h ({\bf r})$ in Eq.(3.7) is the fluctuation part
\begin{equation}
\delta {\bf A}^h ({\bf r}) =\frac 1  2 \int d^2{\bf  r}' \frac {\hat{\bf z}
\times ( {\bf r}- {\bf r}')}
{ | {\bf r}- {\bf r}' |^2}[\rho _h ({\bf r}')- \bar { \rho _h }] ,
\end{equation}
which can be treated perturbatively. We emphasize that in an
anyon  problem, $\delta{\bf A}^h$ would correlate with  anyon density
fluctuation
and represent\cite{laughlin2,chen} a long-range interaction among anyons. Here
$\delta{\bf A}^h$ is
determined by the density fluctuation of {\em holons} instead of spinons
themselves, which belong to an independent
degree of freedom. Thus one expects  $\delta {\bf A}^h({\bf r})$  to provide an
independent dynamical scattering source in Eq.(3.5) just like phonon in a usual
electron system.

The separation (3.7) is meaningful  when the hole
density is not too low. With the presence of a fictitious magnetic field $B_h$,
a new length scale is introduced to the spinon system, which  is the
magnetic cyclotron length
\begin{equation}
 l_c=\frac 1 {\sqrt { B_h }}=\frac a {\sqrt {\pi \delta }}  .\end{equation}
$l_c$  will later be connected to the antiferromagnetic correlation length.
At this mean-field level, a Landau-level structure appears in the spinon energy
spectrum, and a basic energy scale is the cyclotron energy
\begin{equation}
\omega _c=\frac {B_h} {m_s} =2\pi \delta J_s  . \end{equation}
Another energy scale measuring the broadening $\Gamma_s$ of each Landau
level caused by the fluctuation  $\delta {\bf A}^h $
should be also correlated with the doping
concentration. $\Gamma _s $ will be related to an important low-energy scale
in spin dynamics. It can be estimated to be in an order of magnitude
$\sim \delta J$ if the energy scale of the  fluctuating $\delta {\bf A}^h $
is sufficiently small. $\Gamma_s $ could be even sharper when the
fluctuating $\delta {\bf A}^h $ has a higher energy scale as holons become more
mobile at finite doping. Of course, different Landau
levels generally could have different broadening widths, and a further
discussion will be given later.
Similar  broadening problem in a semion system has been recently studied by
Levy  and Laughlin,\cite{levy} where the dynamics of  $\delta {\bf A}^h$ is
already known from the RPA calculation. In the present case, the dynamics of
$\delta {\bf A}^h$ is directly related to that of $\delta
\rho_h=\rho_h-{\bar \rho}_h$, which will be in turn determined
by coupling with
the spinon degree of freedom. So a self-consistent treatment of
the Landau level broadening could be much complicated here.  Nevertheless,
for the purpose of understanding basic
characteristics for spin dynamics, only a general  shape of the Landau level
broadening is needed, and the detailed structure will not be
crucial.

Besides the length and energy scales, we point out that there also exists a
basic temperature scale which is  related to  the
Bose condensation temperature $T_c ^{*} $ of the bosonic spinon  $b_{i \sigma
}$. Recall that for a 2D free boson gas, a 2D Bose condensation is always
suppressed by the thermal excitations at any finite temperature, due to finite
density of states at low energy. In the present case, due to the broadening of
the lowest-Landau-level (LLL), the low-energy density of states presumably will
fall off continuously to zero at the LLL energy bottom.  Thus the low-lying
thermal excitation has vanishing weight at the low-energy  tail, which
could not kill  the Bose-condensation of spinons at a sufficient low
temperature.
As will be demonstrated later, $T_c^*$ will represent an important temperature
scale in spin dynamics.

\subsubsection{Spin susceptibility function }

The spin dynamic structure factor ${S}({\bf q},\omega)$ is defined as a
Fourier transformation  of the spin-spin correlation function
\begin{equation}
S_{\alpha \beta}({\bf q},\omega)=\frac 1 {2\pi} \int_{-\infty}^{\infty}dt
e^{i\omega t} \int d^2 {\bf r}e^{-i {\bf q \cdot r}} < {\bf S}_{\alpha}({\bf
r}, t)\cdot {\bf S }_{\beta}(0,0)>.     \end{equation}
${S}_{\alpha\beta}({\bf q},\omega)$ can be directly measured in
neutron-scattering
experiment, and is related to the dynamic spin susceptibility function
 $\chi _{\alpha
\beta}({\bf q},\omega)$ through the so-called fluctuation-dissipation theorem
\begin{equation}
S_{\alpha \beta}({\bf q},\omega)=\frac 1 {\pi}
\left[1-e^{-\beta \omega}\right]^{-1}\chi_{\alpha\beta} ''({\bf q},\omega),
 \end{equation}
where $\beta=1/k_BT$, and  $\chi ''_{\alpha \beta} (q, \omega) \equiv Im
\chi_{\alpha \beta} ({\bf q}, \omega +i0^+)$ is the imaginary part of the
retarded spin susceptibility function.

In the Matsubara representation, the transverse spin susceptibility is defined
by
\begin{equation} \chi({\bf q},i\omega_n)= \int_{0}^{\beta}d \tau
e^{i\omega _n\tau} \int d^2 {\bf r}e^{-i {\bf q \cdot r}}
<T_{\tau} {S}^+({\bf
r}, \tau) S^-(0,0)>  ,   \end{equation}
with $\omega_n={2\pi n}/ {\beta} $. In terms of a continuum version of
Eq.(2.46), one may
write
\begin{eqnarray}
<T_{\tau} {S}^+({\bf
r}, \tau) S^-(0,0)>=&&\frac {a^2} {4}\left(\sum_{{\bf Q}_0}  e^{i {\bf Q}_0
\cdot{\bf
r}}\right)
<T_{\tau} b_{\uparrow}^+({\bf r}, \tau)  b_{\uparrow}(0,0) e^{i \int _{(0,0)}
^{({\bf r}, \tau)} {\bf A}^h \cdot d {\bf r} } > \nonumber \\
&&\times
<T_{\tau} b_{\downarrow}({\bf r}, \tau)  b_{\downarrow}^+(0,0)
e^{-i \int ^{(0,0)}
_{({\bf r}, \tau)} {\bf A}^h \cdot d {\bf r} } >,\end{eqnarray}
in which ${\bf Q}_0=(\pm\frac {\pi} a, \pm \frac {\pi} a)$ are the AF wave
vectors. $ {\bf   A}^h$ in Eq.(3.14) is from the nonlocal phase in the
spin expression (2.46). In the mean-field approximation with ${\bf   A}^h$
replaced by
$ {\bar {\bf   A}}^h$,  Eq.(3.14) may be rewritten as
\begin{equation}
<T_{\tau} S^+({\bf
r}, \tau)  S^-(0,0)>=\frac {a^2} 4 \left(\sum_{{\bf Q}_0} e^{i {\bf
Q}_0 \cdot {\bf r}}\right) e^{2 i \int _{0}^{\bf r} \bar{\bf A}^h \cdot d
{\bf r} } G_b^{\uparrow}(-{\bf r}, -\tau) G_b^{\downarrow}({\bf r}, \tau),
        \end{equation}
where the line integration on the right-hand side is chosen along a
straight line
connecting $(x,y)$ and $(0,0)$ on 2D plane. The Green's function $G_b^{\sigma}$
is defined by
\begin{equation}
G^{\sigma}_b({\bf r}, \tau )=-<T_{\tau} b_{\sigma}({\bf r}, \tau)
b_{\sigma}^+(0,0) > .  \end{equation}
$b_{\sigma}({\bf r})$ may be expressed in the representation of the Landau
levels as
$b_{\sigma}({\bf r})=\sum _{nk} <{\bf r}|nk> b_{nk}^{\sigma}$, where $n=0,1,2
\cdots , $ is the Landau level index and $k$ is the quantum number inside each
Landau level. Correspondingly,
\begin{eqnarray}
G^{\sigma}({\bf r}, \tau )&&=-\sum_{n,k}\sum_{n',k'}<{\bf r}|nk><n'k'|{\bf 0}>
 <T_{\tau} b^{\sigma}_{nk} (\tau )  {b^{\sigma}_{n'k'}}^+(0) >
\nonumber      \\
&&=-\sum_n <{\bf r}|\left(\sum_k |nk><nk|\right) |0>
<T_{\tau} b^{\sigma}_{nk} (\tau )  {b^{\sigma}_{nk}}^+(0) > \nonumber        \\
&&=\sum_n \Pi_n^{\sigma} ({\bf r},{\bf 0})G_b^{\sigma} (n, \tau).
\end{eqnarray}
In obtaining last line of  Eq.(3.17), the $k$-dependence of the Green's
 function
$G_b=-< T_{\tau} b_{nk}(\tau) b_{nk}^+(0)>$ has been neglected.
Note that $ k$ represents the center of each cyclotron orbital in the present
symmetric gauge and $k$-dependence of $G_b$ should not be important due to the
translational invariance of the system. $\Pi_n^{\sigma} ({\bf r},{\bf 0})$ in
Eq.(3.17) is given by\cite{laughlin2}
\begin{equation}
\Pi_n^{\sigma} ({\bf r},0)=L_n \left(\frac {r^2} {2l_c ^2} \right) \Pi _0 ^{
\sigma}({\bf r},{\bf 0})   \end{equation}
with $L_n(t) $ as the Laguerre polynomials: $L_0(t)=1$, $L_1(t)=1-t$, etc., and
\begin{equation}
 \Pi_0^{\sigma} ({\bf {r}},{\bf 0}) =\frac 1 {2\pi l_c ^2} exp \left[
- r^2/{4 l_c ^2}\right] .  \end{equation}
Then one gets
\begin{equation}
\chi ({\bf q},i\omega_n )=\sum_{l,m=0}^{\infty} K_{lm}({\bf q})\int
_{-\infty}^{\infty} \frac {d \omega' d\omega ''} {2\pi  } \rho _b (l,
\omega ') \rho _b (m, \omega '') \frac { n(\omega '') -n( \omega ' ) }
{i\omega_n +\omega ' - \omega '' }, \end{equation}
where
\begin{eqnarray}
K_{lm} ({\bf q})&&= \frac {a^2} {32 \pi ^3 l_c^4 } \sum _{{\bf Q}_0} \int d^2{
\bf r} L_l
\left(\frac {r^2} {2l_c^2}\right) L_m\left(\frac {r^2} {2l_c^2}\right) e^{-
\frac {r^2} {2 l_c ^2 }}
e^{ -i {\bf r} \cdot ({\bf q}-{\bf Q}_0) } \nonumber   \\
&&=\frac {\delta} {16 \pi } \sum _{{\bf Q}_0}\int _0^{\infty} dy
 L_l (y) L_m (y)e^{-y}
J_0 ( \sqrt {2l_c^2 y }\left|{\bf q} -{\bf Q}_0\right|) ,    \end{eqnarray}
with $J_0(x)$ as the Bessel function, and the spectral function $\rho _b$ is
defined through
\begin{equation}
G_b^{\sigma} (l, i\omega_n)= \int \frac { d\omega '}{2\pi} \frac {\rho_b
(l,\omega ')} { i \omega _n -\omega '} ,   \end{equation}
in which the spin index $\sigma $ has been omitted for simplicity since
$\rho_b$ does not explicitly depend on it in the unpolarized case.
$n(\omega)=1/(e^{\beta \omega}-1)$ in Eq.(3.20) is the Bose distribution
function. Note that the spectral function $\rho_b(l, \omega)$ ensures $\omega
\geq \omega_0-\mu \geq 0$, where $\omega_0$ is the energy minimum of spinon
spectrum  and $\mu$ is the chemical potentail determined by $\sum_l (2\pi
l_c^2)^{-1}\int (d\omega/2\pi)\rho(l, \omega)n(\omega)=(1-\delta)/2a^2$.

Finally $\chi '' ({\bf q}, \omega)$ is found by
\begin{equation} \chi'' ({\bf q}, \omega)=-\pi \sum _{l,l'}K_{ll'}
({\bf q})\int
_{-\infty}^{\infty} \frac {d \omega '} {2 \pi} \rho _b (l, \omega ')
\rho _b (l', \omega +\omega ')\left[n(\omega+\omega ' )- n(\omega ' )\right] .
\end{equation}
 In the next sections, we shall examine the ${\bf
q}$-dependence, ${ \omega}$-dependence and temperature-dependence of $\chi ''
({\bf q}, \omega)$
and compare them with neutron-scattering and NMR experimental measurements in
cuprates.

\subsubsection{Basic characteristics of spin dynamics in 2D}

Spin susceptibility $\chi '' ({\bf q}, \omega)$ in Eq.(3.23) will characterize
the basic features of  spin dynamics. The real part of spin susceptibility can
be also determined  from $\chi '' ({\bf q}, \omega)$ through the Kramers-Kronig
relation. Here we shall mainly  focus on $\chi '' ({\bf q}, \omega)$  in the
vicinity of the AF wave vector  ${\bf Q}_0$ and  low energy $\omega$.

As discussed in Sec.{\em 1}, doping creates a Landau-level structure
in the spinon spectrum. So the low-lying spin fluctuations are expected to be
sensitive to doping. Due to the particular Landau-level structure, one can
distinguish two types of contributions to  $\chi '' ({\bf q}, \omega)$, which
correspond to intra-Landau-level transition and inter-Landau-level transition,
respectively. We will show below that these two processes are  related
to the so-called commensurate and incommensurate AF spin fluctuations in the
present state.

{\em Commensurate AF fluctuation} \hspace {0.3in} First we consider $\chi ''
({\bf q}, \omega)$
at small $\omega$ such that only the intra Landau-level
transition contributes. At low temperature with  spinons staying in the LLL,
the ${\bf q}$ dependence of $\chi'' ({\bf q} , \omega)$ [Eq.(3.23)] is solely
decided by $K_{00} ({\bf q})$:
\begin{equation}
K_{00}({\bf q})=\frac {\delta} {16\pi} \sum _{{\bf Q}_0} \exp \left({-\frac {|
{\bf q} -{\bf Q}_0|^2} { 2
{l_c}^{-2}} }\right) ,    \end{equation}
in terms of Eq.(3.21). Eq.(3.24) shows that  $\chi '' ({\bf q},\omega) $
will be
peaked at the AF wave vector ${\bf Q}_0$'s.  The corresponding
spin fluctuation is known as the commensurate AF fluctuation. With the
increase
of temperature, spinons can be thermally excited to higher Landau levels such
that $K_{11}({\bf q})$, $K_{22}({\bf q})$, etc., will contribute.
For example, $K_{11}({\bf q})$ has the following form:
\begin{equation}
K_{11}({\bf q})=\frac {\delta} {16\pi}\sum _{{\bf Q}_0} \left(1-\frac {|
{\bf q} -{\bf Q}_0|^2}{2{l_c}^{-2}}\right)^2
\exp \left({-\frac {|
{\bf q} -{\bf Q}_0|^2} { 2
{l_c}^{-2}} }\right), \end{equation}
which is still peaked at ${\bf Q}_0$ with a width essentially the same as
$K_{00}({\bf q})$ in Eq.(3.24). $K_{ll}$ at higher $l$ can be checked by
numerical calculation and generally a Gaussian (3.24) is  well satisfied when
$l$ is not too big. This means that the width  is not sensitive to temperature.

Thus, when the intra-level transition dominates, $\chi '' ({\bf q},\omega) $
generally follows a Gaussian:
\begin{equation}
\chi''({\bf q}, \omega)\simeq  \exp \left({-\frac {|{\bf q} -{\bf
Q}_0|^2} { (2
{\sigma}^{2})} }\right) A(\omega),    \end{equation}
around ${\bf Q}_0$,  where $\sigma=1/l_c$ and $A(\omega)\equiv\chi ''({\bf Q}_0
,\omega)$.   The Gaussian (3.26) with a width $\sigma$ determines a spin-spin
correlation in real space: $\cos ({\bf Q}_0\cdot{\bf r})\exp [ -|{\bf r}|^2/\xi
^2]$, where the correlation length
\begin{equation}
\xi=\frac{\sqrt{2}}{\sigma} = a\sqrt{\frac{2}{\pi\delta}} \end{equation}
is in the same order of the average hole-hole distance.  Namely, the doped
holes break up the long-range AF correlation into short-range AF fragments with
a length scale  $\sim \xi$.

The energy scale of this AF fluctuation will be  characterized by the $\omega$
dependence of $A(\omega)=\chi''({\bf
Q}_0, \omega)$ in Eq.(3.26).  Since only the intra-level transition is involved
here, the magnetic energy scale  will be basically decided by the broadenings
of the Landau levels. The expression of $\chi''({\bf Q}_0, \omega)$ can be
found as
\begin{equation}
\chi '' ({\bf Q}_0, \omega)=-\frac {\delta} {16} \int \frac {d \omega ' }
{2 \pi}\sum_{l} \rho_b (l, \omega ')  \rho_b (l, \omega +\omega
')\left[n(\omega+\omega ')-n(\omega ')\right] .  \end{equation}
Here the broadening width $\Gamma_s$ of the spectral function $\rho_b$
for a given Landau level is generally of the order $\delta J$, as caused by
the  fluctuation $\delta {\bf A}_h$. The detailed broadenings will be sensitive
to $\delta {\bf A}_h$ and other factors, which is to
be further discussed later. Equation (3.28) has been  obtained
from Eq.(3.23) by using
\begin{equation}
K_{ll'}({\bf Q}_0 )= \frac {\delta} {16\pi} \delta _{l, \hspace{0.01in} l'},
 \end{equation}
in terms of Eq.(3.21) and the orthogonality of the Laguerre functions
\begin{equation} \int_{0}^{\infty} dy
L_l(y)L_{l'}(y)e^{-y}=\delta_{l,\hspace{0.01in} l'}.
\end{equation}
[The contribution to $K_{ll'}({\bf Q}_0)$ from the  other
three ${\bf Q}_0 $'s is exponentially small and  has  been neglected.]

The typical  $\omega$ dependence of $\chi''({\bf Q}_0, \omega)$ is shown in
Fig.1 at various temperatures.
The choice of the spectral function $\rho_b$ and other parameters in Fig.1 is
to describe the underdoped $YBa_2Cu_3O_{6.6}$, which  will be
explained in the next section. Here we mainly focus on the general features
shown in Fig.1. Besides an overall small energy scale decided by the
Landau-level
broadening, Fig.1 shows an interesting temperature effect
characterized by the Bose-condensation temperature $T_c^*$.  When $T>T_c^*$,
one finds $\chi ''({\bf Q}_0, \omega)\propto \omega/T$ at small $\omega$.  The
slope of the linear
$\omega$ dependence {\it increase} with the decrease of $T$. However, this
{\it increase} gets arrested at $T=T_c^*$, and when $T<T_c^*$ the low-energy
part of $\chi ''({\bf Q}_0, \omega)$ becomes continuously suppressed instead.
This feature resembles a typical ``spin gap'' behavior. When
$T<T_c^{*}$, the spinons begin to condensate into the bottom state of the
lowest Landau level. The contribution due to the transition
from such  a condensate
state to the rest of the quantum states in the broadened LLL will then emerge
in $\chi '' ({\bf Q}_0, \omega)$. This process effectively will map out the
shape of the spectral function of the same Landau level, and becomes dominant
with the decrease of temperature. In fact, at  $T=0$ when all the spinons
are condensed,  $\chi '' ({\bf Q}_0, \omega)$ in Eq.(3.28) is simply reduced
to
\begin{equation}
\chi '' ({\bf Q}_0, \omega)_{T=0}=\left(\frac {n_s}
{16}\right)\rho_b(l=0, \omega)_{T=0} ,\end{equation}
with $n_s=1-\delta$, which is directly proportional to the spectral function
$\rho_b (0, \omega)$ of the LLL. If $\rho_b(0, \omega)$ has
a small Lifshitz tail before it vanishes at the low-energy end  $\omega=0$,
a gap feature is exhibited in $\chi ''({\bf Q}_0, \omega)$ as illustrated in
Fig.1. Even in the
case that $\rho_b(0, \omega)$ does not approach to zero very fast  at
$\omega=0$, $\chi ''({\bf Q}_0, \omega)$ could still generically show a
pseudo-gap trend below $T<T_c^*$ as described above. This spin gap behavior
and the temperature scale $T_c^*$ are the  unique, and important features of
the present spin state.

Another interesting property of $\chi''({\bf q}, \omega)$ is its behavior
at $\omega \rightarrow 0$ limit. This behavior can be probed by NMR
measurement.\cite{nmr} The NMR spin-lattice relaxation rate of nuclear spin
due to the
coupling to spin   fluctuation  described by $\chi''({\bf q}, \omega)$ is
given by\cite{moriya}
\begin{equation}\left.\frac 1 T_1=\frac {k_B T } N \sum _{\bf q }A^2 ({\bf q})
\frac
{\chi ''
 ({\bf q}, \omega
)}{\omega } \right |_{\omega \rightarrow 0} , \end{equation}
where the form factor $A^2({\bf q})$ is from the hyperfine coupling between
nuclear spin and the spin fluctuation. For $\ ^{63}Cu (2)$
nuclear spin in the cuprates,  with the applied field perpendicular  to the
$CuO_2$ plane, the form factor $A^2({\bf q})$ is found to
be\cite{mila,shastry,mmp}
\begin{equation}\left. A^2 ({\bf q})\right|_{^{63}Cu}=\left[A_{\bot}+2B(\cos
q_x a+
\cos q_y
a )\right]^2  , \end{equation}
where the hyperfine couplings $A_{\bot }$ and $B $ are estimated as\cite{mmp}
$A_{\bot
}/B \simeq 0.84 $,
$ B\simeq 3.8 \times 10^{-4} meV$ [These coefficients may slightly vary among
$YBCO$ and $LSCO$ compounds]. For $\ ^{17}O(2)$ nuclear spin, one
has\cite{mila,shastry,mmp}
\begin{equation}
 \left. A^2 ({\bf q})\right |_{^{17}O}=2C^2\left[1+cos (q_x a)\right]  ,
\end{equation}
with $C \simeq 0.87 B $ [Ref.\onlinecite{mmp}]. $A^2 ({\bf q})|_{^{17}O}$ in
Eq.(3.34)
vanishes at ${\bf q}={\bf Q}_0$. Thus a combined measurement of $1/^{63}T_1$
and $1/^{17}T_1$ can provide  a ${\bf q}$-dependent information about $\chi
'' ({\bf q}, \omega )$ at $\omega \rightarrow 0$.

In the present framework, $\chi '' ({\bf q}, \omega )$ have been already
obtained in Eq.(3.23).  So by substituting it into the above $1/T_1$
expression, it is  straightforward to get
\begin{equation}
\frac{1}{^{63}T_1}=\pi \sum_l C_l \int \frac {d \omega  }{2 \pi } n(\omega)
(1+n(\omega)) \left[\rho _b (l, \omega )\right]^2, \end{equation}
where
\begin{eqnarray}
C_l=&&\frac {\delta ^2 B^2 }{8 \pi } \left[\left(\frac{A_{\bot } ^2}{B^2}
+4\right)-8\frac{A_{\bot }}{B}L_l ^2
\left(\frac {a^2 } {\xi^2} \right) e^{-{a^2}/\xi^2 }\right.\nonumber \\
&&\left.+8L_l^2 \left(\frac{2a^2}{\xi^2} \right)
e^{-2a^2/\xi^2 } +4 \left(\frac{4a^2}{\xi^2} \right)
e^{-4a^2/\xi^2 }\right]  . \end{eqnarray}
In obtaining Eqs.(3.35) and (3.36) only the intra-level transition (i.e, $l=l '
 $) is involved.  Using the same $\chi '' ({\bf q}, \omega
 )$ whose $\omega$-dependence at ${\bf q}={\bf Q}_0$ is illustrated in Fig.1,
one finds
the corresponding $1/^{63}T_1T$ vs. $T$ as shown in Fig.2. In contrast to a
usual Korringa rule for
a Fermi liquid: $1/T_1T \sim
const.$, Fig.2 clearly depicts a non-Korringa behavior which  falls off like
$\propto 1/T$ at high temperature.  And $1/^{63}T_1T$ is peaked
at the characteristic temperature $T_c^*$, below which   $\chi''({\bf Q}_0,
\omega)$ begins to develop  a ``spin gap'' behavior.

On the other hand, at  oxygen  $\ {^{17}O(2)}$ site, the
coefficient $C_l$ has to be  replaced by $C_l^O$ in Eq.(3.35):
\begin{equation}
C_l^O=\frac {\delta ^2 C^2 }{4 \pi } \left[1-L_l^2 \left(\frac{a^2}{\xi^2}
\right)e^{-{a^2}/{\xi^2 } }  \right]  . \end{equation}
The  difference between $1/^{63}T_1$ and $1/^{17}T_1$ is due to the different
weight functions [Eqs.(3.33) and (3.34)], which pick up the commensurate
magnetic fluctuation around ${\bf q}={\bf Q}_0$ at $^{63}Cu(2)$ site, but
suppresses such a contribution at $^{17}O(2)$ site.
$C_l^O$ in Eq. (3.37) simply vanishes at $\xi \rightarrow \infty$: the
long-range AF limit where the whole contribution is from ${\bf q}={\bf Q}_0$.
At $\xi=4 a$, for example,  we find a
reduction of $\sim 1/93$ for $1/^{17}T_1$  as compared to $1/^{63}T_1$.  Thus
the commensurate AF fluctuation as described by Eq.(3.26) has negligible
contribution to the spin relaxation for planar oxygen nuclei of the cuprates.
In contrast, non-magnetic incoherent  contributions should not be
suppressed  so strongly by the form factor, and  would become dominant
contribution in $1/^{17}T_1$ as well as in  the  Knight shift,\cite{nmr} which
measures the real part of the spin susceptibility near ${\bf q}=0$.

Hence,  $\chi'' ({\bf q}, \omega )/\omega$ at $\omega \rightarrow 0$ gives rise
to  a non-Korringa behavior of the NMR spin relaxation rate. This $1/T$ law of
$1/T_1T$ can  be obtained analytically, if the broadening  $\Gamma_s^0$ of the
spectral function $\rho_b$ for the LLL is small as compared to temperature.
That is, when $\Gamma_s^0\ll T\ll \omega_c$ (and
of course $T>T_c^*$), one has
\begin{equation}
\frac{1}{^{63}T_1T}\simeq \frac{D}{T},  \end{equation}
in terms of Eq.(3.35), where $D=\pi C_0n_s/\Gamma_s^0 \delta^2$.
Next let us  consider the case when temperature is further increased such
that more Landau levels are involved.  Here we have
to assume a general Lorentz-like broadening for each Landau level with
$\Gamma_s=0.4\omega_c\propto \delta$.  Then $1/^{63}T_1$ at high temperature is
shown in
Fig.3, where the curves become very
flat and not sensitive to the doping concentration. This is  in contrast to the
low temperature regime, where a strong doping-dependence can emerge (in
Eq.(3.38), $D\sim 1/\delta$ if one simply takes $\Gamma_s^0\propto \delta$).

Therefore,  a full picture for the low-lying commensurate AF correlation
is formed in the present spin-charge separation state. This picture is
rather unique, and is drastically different from those of Fermi liquid
and local antiferromagnetic descriptions. Doping effect plays a key role
here.   It decides a doping-dependent energy
scale for  the AF fluctuation, which  can be much smaller than the
characteristic energies in a Fermi liquid ($\epsilon_f$) and an effective local
antiferromagnet ($[1-\delta]J$).   It also leads to a doping-dependent
correlation length, comparable with the average spacing of holes. Furthermore,
the Bose condensation of spinons determines a new characteristic temperature
scale, below which the low-lying spin fluctuation is suppressed and
non-Korringa behavior of the spin relaxation rate gets interrupted, similar to
a spin gap effect.

{\em Incommensurate AF fluctuation}\hspace{0.3in} Now let us consider higher
energy regime. If $\omega$ is increased such that the inter Landau-level
transition gets involved, the simple Gaussian-like ${\bf q}$-dependence in
Eq.(3.26) will
be modified.  A typical $\chi''({\bf q}, \omega)$ vs. ${\bf q}$ is shown in
Fig.4 at different
$\omega$'s. For the purpose of
illustration, we have chosen the same broadening $\Gamma_s=0.4\omega_c=0.4
\delta
\omega_0$ for
each Landau level, and $T=0.1\omega_0$, $\delta=0.15$.
It shows that the width of the Gaussian is  broadened at first  with the
increase of $\omega$. Then the commensurate peak
at ${\bf Q}_0$ is split and two incommensurate peaks emerge at some fixed
positions as the inter-level transition becomes dominant. In fact, this
incommensurate structure is due to  $K_{01}({\bf q})$ in Eq.(3.23).
According to Eq.(3.21), one finds
\begin{equation}
K_{01}({\bf q})=\frac {\delta} {16\pi}\sum_{{\bf Q}_0} \frac {|
{\bf q} -{\bf Q}_0|^2}{2{\sigma}^{2}}\exp \left({-\frac {|
{\bf q} -{\bf Q}_0|^2} { 2
{\sigma}^{2}} }\right),  \end{equation}
which leads to incommensurate peaks at a ring circling ${\bf Q}_0$ by a
radius of $\sqrt{2\pi\delta}/a$.  We stress that the doping concentration is
presumably small here so that the lattice effect is negligible. At a finite
doping where spin-correlation length becomes comparable with the lattice
spacing, the lattice effect is expected to become important, which could
strongly affect  the positions of the incommensurate peaks
in ${\bf q}$ space. In this case, both flux and lattice have to be treated on
the same footing like in the incommensurate-flux-phase.\cite{cfp}

With the further increase of $\omega$, more complicated structure will show up
as $K_{02}$, $K_{03}$, etc., are involved. The weight of $\chi''({\bf q})$ will
be further shifted towards whole Brillouin zone, i.e., non-magnetic regime. At
such a
high energy, the present continuum approximation may no longer be
appropriate. Nevertheless,
one could still get some  generic feature. By integrating ${\bf q}$
within the whole ${\bf q}$ space, we get
\begin{eqnarray}
\chi ''_{total} (\omega )&&=\frac 1 N \sum _{\bf q }\chi '' ({\bf q}, \omega)
\nonumber \\
&&=\frac {\delta ^2} {8\pi} \int \frac {d \omega ' } {2 \pi } \left[\sum _l
\rho_b
(l, \omega ')\right]  \left[\sum _m \rho_b (m, \omega+\omega ')\right]
(n(\omega +\omega ')
-n(\omega ')) .                              \end{eqnarray}
Equation (3.40) predicts that at high energy, the  ${\bf q}$-integrated
$\chi''_{total}(\omega)$ will be saturated at a constant level,  modulated by a
Landau-level-like oscillation. We note that
the higher energy part is contributed by those fluctuations with less
magnetic character. Thus it is less easy for elastic neutron-scattering to
fully collect data for $\chi''_{total}(\omega)$. On the other hand, at low
energy,
finite widths of the Landau levels will lead to  a more broadened first peak in
$\chi''_{total}(\omega)$  as compared to the overall energy peak of
$\chi''({\bf Q}_0, \omega)$ where only intra-level transitions are involved.

\subsection{Comparison with experimental measurements in cuprates}

The normal-state spin dynamics represents an important characteristics
for  the high-$T_c$ cuprates. A lot of measurements have been done on these
materials, especially $YBCO$ and $LSCO$ compounds.  Experimental results have
revealed a
rather rich phenomenon, and also seem to suggest that
material-dependent effects,
like double-layer structure in $YBCO$, may play major roles in the spin
dynamics. Thus, whether there is an underlying universal mechanism for
spin dynamics in the cuprates is not transparent  in terms of  the
experiments alone. In the following, we argue that the present spin-charge
separation scheme can relate a variety of anomalous spin properties together,
and provide  a consistent picture for these materials.

In the underdoped cuprates, like $YBa_2Cu_3O_{6.6}$ and insulating
$La_{1.95}Ba_{0.5}CuO_4$,
a commensurate AF fluctuation has been verified by  neutron
scattering,\cite{ybco1,ybco2,hayden} which is  peaked at
${\bf Q}_0$ in $\chi''$. These neutron data have been well
fitted\cite{ybco2,hayden}
by a Gaussian form (3.26), and the width $\sigma$ is indeed roughly independent
of
temperature. The corresponding spin-spin correlation length follows a
$a/\sqrt{\delta}$
rule similar to Eq.(3.27), as shown\cite{lsco} in $LSCO$ at small doping.  Thus
the momentum feature of the commensurate AF fluctuation at small doping is well
described by the theory.
In the  underdoped $YBa_2Cu_3O_{6+x}$, the energy scale of such a commensurate
AF
fluctuation has been systematically investigated.\cite{ybco1,bourges,chou}
It is found to be doping dependent and small as compared to  the
exchange energy
$J\simeq 120$ $meV$. As discussed in the last section, a small,
doping-sensitive
energy scale for the commensurate AF fluctuation
is one of intrinsic features of the present scheme, and is distinguished
from the usual Fermi liquid as well as the local spin descriptions. To our
knowledge,
so far there is no other alternative theory could be able to obtain such a
small energy scale ($\ll J$) for the commensurate AF fluctuation.
Furthermore, a prominent feature has been exhibited\cite{ybco1,ybco2} at the
low energy
part of the spectroscopy,
where the  weight of $\chi''$ is continuously suppressed below some
characteristic temperature, resembling an opening of a spin gap. In the present
theory, even though there is no real gap in the spin fluctuation spectrum,
a characteristic temperature $T_c^*$ is found to be naturally associated with
a spin-gap-like phenomenon. In Fig. 1, a
typical energy and temperature dependence of $\chi ''({\bf Q}_0, \omega)$ has
been shown, where the spectral function is chosen to describe
$YBa_2Cu_3O_{6.6}$
(see below). And the overall energy and temperature features in Fig.1,
particularly
the ``spin gap'' behavior, are in good agreement with
those found in the underdoped  $YBa_2Cu_3O_{6+x}$.\cite{ybco1,ybco2}
In the underdoped cuprates, the NMR spin relaxation rate of planar copper
nuclei\cite{takigawa}
manifests a non-Korringa behavior, which is
suppressed when the spin gap feature shows up in the neutron scattering
at low temperature. In contrast, a conventional
Korringa temperature behavior\cite{takigawa} is found for  planar oxygen
nuclei. The
theoretical  $1/^{63}T_1T$ has been presented in Fig.2, which is
calculated by using the same spectral function as used in Fig. 1 for
$YBa_2Cu_3O_{6+x}$.  Fig. 2 shows a non-Korringa behavior at $T>T_c^*$ as well
as a
``spin gap'' feature below $T_c^*$. All of them are also qualitatively
consistent  with the
experimental measurements [the contribution of the present {\em commensurate}
AF
fluctuation to
$1/ ^{17}T_1T$ is dramatically suppressed ($\sim 1/93$ at $\xi \sim 4a$ for
$YBa_2Cu_3O_{6.6}$)
so that the non-Korringa signal does not leak to the oxygen sites].
Therefore,  the ${\bf q}$-, $\omega$- and $T$- dependence of the present
low-lying magnetic fluctuation at small doping are all in agreement with the
main experimental features found in the underdoped cuprates.

Techniquely, let us discuss the spectral function used in the theoretical
calculation. According to
Eq.(3.31), one may use the experimental measurement of $\chi ''({\bf Q}_0,
\omega)$  at
low temperature to
determine $\rho_b(0, \omega)$, instead of a first-principal calculation
which would
involve much complicate factors here like inter-layer coupling. As $\rho_b(0,
\omega)$ satisfies the normalized condition, if its broadening at low
temperature is
quite large as compared to the temperature scale which we are
interested, the temperature dependence of $\rho_b(0, \omega)$ may not be
important. Then $\chi ''({\bf Q}_0, \omega)$ in  a whole
temperature range can be determined. In Fig.1, we have chosen $\rho_b(0,
\omega)$ such that to give a right energy
scale of $\chi ''({\bf Q}_0, \omega)$ for $YBa_2Cu_3O_{6.6}$
at low-temperature [Fig.7 in Ref.\onlinecite{ybco2}].
The contribution from the higher Landau levels are neglected because
the temperature range considered is comparatively smaller than the
broadening of the LLL. Besides the shape of $\rho_b(0,
\omega)$, an overall strength of $\chi ''({\bf Q}_0, \omega)$ has   also been
adjusted,
in order to compare with experiment, by reducing the spinon concentration from
$n_s=1-\delta$ to $n_s^*<n_s$. This is because in the present approximation the
spinons are treated as ideal
bosons instead of hard-core bosons. At
half-filling, this approximation could lead to a too big magnetization. In the
present
doped case, it would also give rise to  a too strong magnetic
fluctuation. The hard-core effect as well as the incoherent band
in the spinon spectrum should  reduce the effective number of spinons
contributing to the AF correlation. We find a reduction of $n_s^*/n_s\sim
1/3$ at $\delta=0.10$ giving  a $T_c^*\simeq 175 K$, in consistence  with the
corresponding  experimental characteristic temperature ($\sim 160$
$K$),\cite{ybco1,ybco2} and {\it at the same time}, leading to a spin
relaxation rate whose magnitude
quantitatively agreeing with the NMR measurement.\cite{takigawa,nmr}

To end  discussion of the underdoped materials, we give several remarks below.
Some authors\cite{millis,altshuler,ubbens} have
attributed the spin gap phenomenon in the underdoped $YBCO$ to their peculiar
bi-layer structure. In contrast, one may have noticed that the theory here is
purely two dimensional. Nevertheless, it has been  noted that the spinon
spectral function $\rho_b(0, \omega)$ for the LLL has been determined directly
from the
experiments, in which the inter-layer coupling could have been already included
and may play a key
role in the broadening.
Recall that in the present scheme, there is no real gap opened in the spin
spectrum, and
the ``gap'' feature is very sensitive to the detailed broadening
of the LLL. Theoretically, for a pure 2D system one would expect the broadening
of the LLL to be much narrower at low temperature. The  reason is that a spinon
in the LLL could not be scattered into a
lower energy state while emits a phonon-like excitation of $\delta {\bf A}^h$,
because it is already at the energy bottom. It cannot jump up to higher
Landau-levels either, due to the energy conservation.  Such a feature has been
indeed
found  in a similar problem.\cite{levy} If this were true,  the energy
scale in the underdoped material might  have been much sharper than observed.
But in the insulating cuprates like
$La_{1.95}Ba_{0.5}CuO_4$ [Ref.\onlinecite{hayden}],  holons should be
localized and
randomly distributed
such that $\delta {\bf A}^h$ describes a random flux with energy scale
$\simeq 0$, instead of a phonon-like dynamic mode. Then the
degeneracy of the LLL can be lifted under the  static random vector potential,
and consequently a much broadened LLL could appear. And mixing with other
levels is also expected here. For the underdoped $YBCO$, however, one is in the
metallic phase, and holes must be mobile which would lead to a well-defined
dynamics for $\delta {\bf A}^h$. Nonetheless, one may still expect a strong
localization tendency
of holes at low temperature, as suggested by the transport
measurements,\cite{res} which could in turn lead to a more broadened LLL
than in the
optimally-doped regime. {\em Furthermore}, the bi-layer coupling in $YBCO$
can split
a sharp LLL into two peaks (symmetric and antisymmetric states), but for  a
larger broadened LLL, the bi-layer coupling may well result in a single, much
broadened peak like the one shown\cite{ybco2} in $YBa_2Cu_3O_{6.6}$. Finally,
we point
out that in these underdoped materials,
incommensurate structure has not been observed yet at high energy. But the
width $\sigma$ in Eq.(3.26) has been found to be increased with the energy,
which is an indication of  the involvement of inter-level
transition in the present theory.

Next we consider optimally doped cuprates, like $YBa_2Cu_3O_{7-y}$ and
metallic $La_{2-x}Sr_xCuO_4$. A striking common feature in these materials is
the {\em lack} of a commensurate AF fluctuation at low energy in  neutron
measurements. The spin-polarized neutron-scattering measurement\cite{mook} of
$YBa_2Cu_3O_{7-y}$ has only revealed a sharp  high energy AF peak around
$41\hspace{0.05in}meV$. In the metallic
$LSCO$ ($x=0.075$, $0.14$ and $0.15$), an incommensurate structure has been
found\cite{inc}
down to an energy scale $\simeq 1 \hspace{0.1in} meV$, and no
commensurate
AF correlation is observed within the experimental resolution.
However, the absence of a low-energy AF fluctuation or the presence
of a low-energy incommensurate
fluctuation would be both  in sharp conflict with the  NMR
measurements\cite{nmr}
which probes
$\omega \simeq 0$ regime and implies a strong {\it commensurate} AF
correlations in these
materials.
For example, by extrapolating the spin susceptibility with the
incommensurate structure observed by neutron-scattering
down to the  NMR frequency ($\sim 10^{-4}\hspace{0.05in}meV$), it has  been
found\cite{walstedt,barzykin} that a
large magnetic contribution could leak to $^{17}O(2)$ sites, in contrast
to  the measured spin relaxation rate $1/{^{17}T_1}$ which appears\cite{nmr} to
be
completely
dominated by a nonmagnetic ($\sim {\bf q}$ independent) contribution.  As
a matter
of  fact, the spin relaxation rates in $LSCO$ have shown canonical
behaviors, which
basically are the same as those found in $YBCO$ compounds.
As elaborated in Sec.{\em B3}, only a
commensurate contribution can lead to a non-Korringa behavior of $1/T_1$ at
$^{63}Cu(2)$ sites, and {\it at the same time},  be strongly cancelled out at
$^{17}O(2)$ sites. This fact has been the basis for the so-called nearly
antiferromagnetic-Fermi-liquid theory.\cite{mmp,mbp}  Therefore, in order to
reconcile neutron and NMR
experiments, one is led to the conclusion that a commensurate AF fluctuation
should re-emerge in the optimally-doped cuprates within an energy scale beyond
the
experimental resolution.

However, it is hard to perceive such a characteristic
scale in a conventional theory.
So far, there have been a number of theoretical conjectures for the mechanism
of
incommensurate magnetic fluctuations in the metallic  $LSCO$. In the
Fermi-liquid-like
framework, the
incommensurability is directly connected\cite{fl,littlewood} with the
Fermi-surface
shape. But it
lacks a small energy scale, within which a commensurate structure could be
recovered. The spiral state of Shraiman and Siggia\cite{ss2}  based on the
$t-J$ model
also provides an incommensurate structure. The original spiral state is a
long-ranged state, and some short-range versions have been
proposed.\cite{kane,w}
But a low-energy commensurability is still hard to be comprehended here.
Some non-intrinsic mechanism for the incommensurability due to
the inhomogeneity in the $LSCO$ system has been also proposed\cite{barzykin} in
order to
reconcile the neutron and NMR data, where the commensurate AF fluctuation
is assumed intrinsic in the metallic $LSCO$.  Nevertheless, the details still
need to be carried out in order to make comparison with the experiments.

The present spin-charge separation scheme is unique to have a small
characteristic energy scale of the commensurate AF fluctuation,
as already discussed in the underdoped case. As pointed out there, an even
smaller energy scale could be present at optimal regime as the LLL broadening
becomes sharper when holons are much mobile in this larger doping regime.
Due to the narrowness of the energy range,
the amplitude of $\chi''$  will be also small due to the cancellation of the
Bose
functions inside Eq.(3.23) [one has
$\chi''\rightarrow 0$ at $\omega/T\rightarrow 0$]. Therefore, this commensurate
AF
correlation may well be beyond the experimental  resolution to be
directly observed by neutron-scattering. In $YBa_2Cu_3O_{7-y}$, the bi-layer
coupling could also  split the sharp LLL into two peaks as mentioned before. It
then explains the sharp $41\hspace{0.05in}meV$ AF peak found by
neutron-scattering\cite{mook} as a result of the spinon transition between
these two
split peaks. On the other hand, such an AF fluctuation can
give rise to a non-Korringa law of the NMR spin relaxation rate [Eq.(3.38)],
which is consistent
with the NMR measurements in $YBCO$ and $LSCO$.  Imai {\em et al.}\cite{imai}
have
measured $1/^{63}T_1$ in $La_{2-x}Sr_xCu_2O_4$ system up to 900K.
At high temperature, all data ($x=0-0.15$)  seem to converge and
saturate to the same temperature-independent value $2700 \pm 150 sec ^{-1}$.
The theoretical results shown in Fig.3 agree well with this tendency. And if
we choose, say, $\omega _0=1000 K$, we find that the  saturation value of
$1/^{63}T_1$ is about twice larger than the measured one, which is a
reasonable
value, considering no  adjustment of the spectral function has been made to fit
the data.

In our theory, the incommensurate fluctuation in the metallic $LSCO$ will be
attributed to the dominance of the inter-Landau-level transition in the
experimental
energy-transfer regime,  as discussed in Sec.{\em B3}. To be
consistent with neutron-scattering measurements, the LLL
broadening has to be very sharp as explained above, while the
second level broadening is relatively larger, which should be centered around
$\omega_c\sim 10 \hspace{0.08 in}meV$ in $x=0.15$ case. When the  temperature
is increased such that $k_BT\sim \omega_c$ , a large amount of
spinons are expected to be thermally excited to the second Landau level. Then
in the
experimental-observable energy regime, the intra-level transition could emerge
again due to a larger
broadening in the second Landau level.
Correspondingly, the incommensurate structure should be
replaced once again by the commensurate peak around ${\bf Q}_0$. This has been
indeed observed in the neutron-scattering  measurement,\cite{inc} where
a broad
commensurate peak is found to re-emerge around $T\sim 100 K$.

The characteristic  energy scale $\omega_c$ of $YBCO$ seems to be
several times larger than  that of $LSCO$ for some unknown reason.  This
decides a main distinction
of spin dynamics between $YBCO$ and $LSCO$, in terms of the present theory.
In other words, one should expect that at a sufficient high energy,
incommensurate structures could also show up in $YBCO$ system. Of course,
many factors, especially the inter-layer coupling, may complicate the details.
Due to these incommensurate
contributions at high energy, the ${\bf q}$-integrated
susceptibility function $\chi''_{total}$ [Eq.(3.40)] will stretch up to an
overall energy
scale $\sim J_s$ with a roughly-constant amplitude (modulated
by the oscillation), Such a unique $\chi''_{total}$ behavior is a high energy
prediction of the present theory for the cuprates.

In the optimally-doped $LSCO$ and $YBCO$, the spin gap
feature is {\em absent} in the commensurate spectrum due to the fact that the
energy scale is too small. But according to the theory, a  pseudo-gap should be
also present in the $LSCO$ compounds below $T_c^*$, which involves the
inter-level transition instead of intra-level transition in the commensurate
fluctuation case. This pseudo-gap phenomenon has been clearly shown by the
neutron  scattering.\cite{inc} In these optimally-doped cases, the
spin characteristic temperature $T_c^*$ should become
very close to the superconducting transition temperature $T_c$, to be
consistent with neutron scattering and NMR measurements. In the present theory,
superconducting
transition will occur when spinons and holons are both condensed. Usually one
finds
$T_c^*>T_c$ at small doping. But when the spinons are condensed, the
frustration effect on holons from
the spin part
(see next section) will be reduced too, which in turn is in favor of the Bose
condensation of holons. In other words, $T_c$ and $T_c^*$ may correlate with
each other.
An optimal regime in the present theory may be properly defined as when
$T_c^*$
and $T_c$ coincides.  A further discussion of the superconducting  transition
will be presented elsewhere.

\section{TRANSPORT PROPERTIES}

Due to spin-charge separation, holon degree of freedom is to be solely
responsible for the electron transport phenomenon in this system.
As noted before, the nonlocal phase ${A}_{ij}^s$ in the holon Hamiltonian
Eq.(1.3) will provide an unconventional scattering mechanism in two dimensions,
as in contrast to the 1D case where it vanishes and holons simply behave like
free particles. In the following,  we will explore 2D transport properties
under such an effective Hamiltonian $H_h$. Similar to the magnetic properties
discussed in Sec. III, transport will be an another crucial test for the
experimental relevance of the present spin-charge separation theory.

For simplicity,  we may consider $H_h$ in the continuum limit at small doping.
The validity of the continuum approximation will be discussed later. The
corresponding continuum version of Eq.(1.3) has the following form\cite{note2}
\begin{equation}
{\tilde H}_h =\int d^2 {\bf r}\ \ h^+ ({\bf r}) \frac {(-i \nabla -
{\bf A}^s)^2}{2m_t} h({\bf r}) ,        \end{equation}
where the holon mass $m_t=({\sqrt 2} t_h a^2)^{-1} $, and ${\bf A}^s $ is
defined by ${
{\bf A}}^s = {{\bf A}}^s_ + +{{\bf A}}^s_- $, with
\begin{equation}
{{\bf A}}^s_{\sigma } =\frac {\sigma } 2 \int d^2{\bf r}' \frac
{\hat{\bf  z} \times ({\bf r-r'})}   { |{\bf r-r'}|^2} \rho _{\sigma }^s (
{\bf r}')  . \end{equation}
Here $\rho _{\sigma }^s ({\bf r}) =b_{\sigma }^+({\bf r})  b_{\sigma }({\bf
r})$ is the local spinon density with spin index $\sigma$. Similar to the
lattice version [Eq.(1.5)], equation (4.2) describes fictitious $\pi$-flux
quanta bound to spinons, and the sign $\sigma=\pm 1$ in front of the
integration means that spinons with different spins carry flux tubes
in opposite directions (see Fig.5). In an unpolarized system, one has
$<\rho_{\uparrow}^s>=<\rho_{\downarrow}^s>$ so that on   average $<{\bf
A}^s>=0$. In other words, the scattering
source as contributed by spinons may be regarded as a sort of fluctuating
gauge field, and no $T$- and $P$-symmetry violations occur here.

Gauge-fluctuation-related scattering mechanisms have been intensively studied
\cite{larkin,lee,ioffe}
within the framework of the uniform RVB state. So we need to distinguish the
present gauge field ${\bf A}^s$ and those studied in the literature.
Since the fictitious flux  quanta are bound to spinons in the present case,
let us first  consider the local density distribution of spinons.
At small doping, spinons are forming
fluctuating AF domains, as discussed in Sec. III, within the spin correlation
length $\xi> a$. For the purpose of illustration, an
extreme case is shown in Fig.6, where $\xi\sim a$ and only a
pair of $\uparrow$ and $\downarrow$ spinons are bound together like a valence
bond. $+$ and $-$ signs in Fig.6 represent a spin configuration, or
equivalently, directions of the $\pi$-flux quanta attached to
spinons. If a holon moves about a closed path $C$ as shown in Fig.6, the
accumulated flux
will obviously have a large cancellation within the loop, with
only the contribution from those near the path $C$ whose pairs are cut by the
loop. Then the mean-square accumulated flux will follow a perimeter law:
$\propto L_C$ ($L_C$ is the perimeter of the loop C).  It is easy to observe
that this perimeter law is generally true  as long as there exists a
short-range
AF correlation, which is present at small doping and within a temperature range
$k_B T<J$. This is in contrast to the gauge fluctuation in the uniform RVB
state where fluctuating flux satisfies\cite{lee,lee3} an area law instead of a
perimeter law.
And the  present flux problem  is more similar to  a random flux problem  in
the so-called ``Meissner phase'',\cite{gavazzi} where a perimeter law is
present.
But in the
conventional case, the strength of the
fluctuating flux is presumably weak  so that the accumulated flux enclosed
in a loop will become negligible when the loop is small enough. In this
case only the long-distance behavior matters, and a
perturbative approach may be applicable which is similar to a usual
phonon-scattering problem.

However, in the present case, each flux tube  attached to a spinon is
quantized at $\pi$.  It means that a slight deformation of the path C, with
one spinon enclosed or excluded, could lead to an additional Berry-phase
$\pm \pi$, or a sign change of the wavefunction.
A  strong  phase interference is therefore  expected at a {\em
short-distance}\cite{note1} due to the  high density of spinons at small
doping. Such a short-distance effect can  drastically change the
nature of scattering  mechanism which in the usual case would only involve
long-wavelength processes. This is going to be a key distinction between the
present theory and the usual gauge theories. We will see that such a
short-range interference can lead to an exotic ``localization'' effect of
holons, and result in a set of very interesting transport anomalous in
longitudinal and transverse channels.

We begin by  considering  a single holon problem. For an one-body problem, a
Feynman path-integral formulation will become very useful. The transition
amplitude
for a holon to travel from  $a$ to $b$ in space under the vector potential
${ {\bf A}}^s$ is given by\cite{feynman}
\begin{equation}
K(b,a)=\int _{a} ^b {\cal D}{\bf r}(t)\hspace{0.02in} e^{i {\cal S} [b,a] }
, \end{equation}
with the action
\begin{equation}
{\cal S} [b,a]=\int _{t_a}^{t_b} dt \left[\frac {m_t } 2 { \dot {\bf r}}^2-
{ \dot {\bf  r}} \cdot ({{\bf A}}^s_+ + {{\bf A}}^s_-)\right].
\end{equation}
By introducing the following identities
\begin{mathletters}
\label{generallabel}
\begin{equation}
\int \frac {d^2{\mbox {\boldmath $\lambda$}}} {(2 \pi )^2 } e^{i {\mbox
{\boldmath $ \lambda $}}\cdot ({\bf r-r}_p )}=\delta ({\bf r-r}_p ),
\end{equation}
\begin{equation}
\int \frac {d^2{\mbox {\boldmath  $\beta$}} } {(2 \pi )^2 } e^{i {\mbox
{\boldmath $\beta $}}\cdot ({\bf r-r}_q )}=\delta ({\bf r-r}_q ),
\end{equation}
\end{mathletters}
$K(b,a) $ may be rewritten as
\begin{equation}
K(b,a)=\int_{b} ^a {\cal D}{\bf r}(t) {\cal D}{\bf r}_p(t) {\cal D}{\bf r}
_q(t)   \int {\cal D}{\mbox {\boldmath $\lambda$ }}(t)   \int {\cal D}
{\mbox{\boldmath $\beta$ }}(t)\hspace{0.02in}
 e^{i {\cal S}_{\lambda \beta} [b,a] }, \end{equation}
in which
\begin{eqnarray}
{\cal S}_{\lambda \beta} [b,a]=&&\int _{t_a}^{t_b} dt \left[\frac {m_h }
 2 {\dot{\bf r}} ^2  + \frac {m_p } 2 { \dot {\bf  r}}_p ^2
 + \frac {m_q } 2 {\dot {\bf  r}}_q ^2 \right]
-\int _a^b d{\bf r}_p  \cdot {{\bf  A}}^s_+({\bf r}_p )
-\int _a^b d{\bf r}_q  \cdot {{\bf A}}^s_-({\bf r}_q ) \nonumber \\
&&+\int _{t_a}^{t_b} dt [{\mbox {\boldmath $\lambda$}}(t)\cdot ({\bf r-r}_p)
+{\mbox {\boldmath  $\beta$}}(t)\cdot ({\bf r-r}_q)] ,      \end{eqnarray}
where $m_h+m_p+m_q=m_t$. Extra degrees of freedom are thus introduced in
Eq.(4.6). Without the last term involving the Lagrangian multipliers ${\mbox
{\boldmath $\lambda$ }}$ and ${\mbox {\boldmath $\beta $}}$, ${\cal S}_
{\lambda \beta} [b,a]$ in Eq.(4.7) would simply describe three
independent particles: a free holon with mass $m_h$ known as $h$ species, $p$
and $q$ species which interact with the vector potentials
${{\bf A}}^s_+ $ and ${{\bf A}}^s_- $,
respectively. The fields ${\mbox
{\boldmath $\lambda$} }$ and ${\mbox {\boldmath $\beta $}}$ play the role
to recombine these three species together as a real holon, and thus effectively
eliminate the additional degrees of freedom in the end.

So far no approximation has been made. By introducing $p$ and $q$ degrees of
freedom, the effects of   ${ \bf A}^ s_+ $ and
${ {\bf A}}^s_- $  are separated. In terms of
Eq.(4.2), ${{\bf A}}^s_{\sigma} $ may be rewritten as ${{\bf
A}}^s_{\sigma} =\sigma {\bar {\bf A}}^s + \delta {\bf
A}^s_{\sigma} $, where
\begin{equation}
{\bar {\bf A}}^s=\frac{B_s}{2} (\hat{\bf z} \times {\bf r}),  \end{equation}
with $B_s=(1-\delta)\pi/2$, and $\delta { \bf A}^s_{\sigma} $ has the
same form as Eq.(4.2) but with $\rho^s_{\sigma}$ replaced by $\delta
\rho^s_{\sigma} = \rho^s_{\sigma} - <\rho^s_{\sigma}>$.  Similar procedure has
been used
in dealing with the topological phase ${\bf A}^h$ in spinon system
(cf. Sec. III), and is familiar in an anyon problem.\cite{laughlin2,chen} Here
$\sigma
{\bar{\bf A}}^s$
describes a mean-field effect with the flux-quanta smeared out uniformly in
space. And $\delta {{\bf A}}^s_{\sigma} $  represents the fluctuation of the
flux-quanta due to the density fluctuation of spinons with spin index
$\sigma$. In the following, we shall show that, to a
leading order of approximation, the effect of  $\delta { {\bf A}}^s_{
\sigma} $ in the action (4.7) could be effectively represented by a relaxation
 of the binding constraint enforced by
${\mbox {\boldmath $\lambda$ }}$ and ${\mbox {\boldmath $\beta $}}$ within a
scale of the cyclotron length $d_c$ of $p$ and $q$ species (under the
fictitious field
$B_s$). Under such an approximation a renormalized
holon will be a composite particle of a bare holon and the $p$ and $q$
species which are in the cyclotron orbitals. Below we elaborate this
approximation.

Due to the symmetry, one only needs   to  focus on the  $p$ species.
Let us consider an arbitrary closed path  $C$ on a $2D$ plane (Fig.7). The
contribution of ${{\bf  A}}^s_{+} $ to the transition amplitude  in
Eq. (4.6) for such a path will be a gauge-invariant  phase:
\begin{equation}
\oint_C d{\bf r}_p\cdot {\bf {{A}}}^s_+= \oint_Cd{\bf r}_p\cdot {\bar
{\bf  A}}^s + \oint_Cd{\bf r}_p\cdot \delta {\bf { A}}^s_+.
\end{equation}
Eq. (4.9) may be further rewritten as
\begin{equation}
\oint_C d{\bf r}_p\cdot {\bf {A}}^s_+= \oint_{C_p}d{\bf r}_p\cdot
{\bar {\bf  A}}^s , \end{equation}
in which a path $C_p$ is introduced such that
\begin{eqnarray}
\left(\oint_{C_p}- \oint_{C}\right)(d{\bf r}_p\cdot {\bf {\bar A}}^s)&&=
\oint_{C}d{\bf r}_p\cdot \delta{{\bf  A}}^s_+ \nonumber \\
&&=\pi \Delta N_c^{\uparrow}. \end{eqnarray}
Here the path $C_p$ as shown in Fig.7 is a deformation of the path $C$ to
account for the fluctuation of $\delta {\bf A}_s^+$ in terms of the
mean-field ${\bar {\bf A}}^s$. In  second line of Eq.(4.11),  $\Delta N_c^
{\uparrow}$ represents the total fluctuating $\uparrow$  spinon number
(with regard to the average one) enclosed by the path $C$, and such a term
describes the total {\em fluctuating } fluxes bound to $\uparrow$ spinons as
deduced from the right-hand-side integral in the first line.
So the  flux enclosed between the paths $ C$ and $C_p$ under
a uniform field $B_s$ will represent the  fluctuation of the flux quanta
attached to $\uparrow$ spinons inside the path $C$. In the following we shall
point out that $\pi\Delta N^{\uparrow}_c$ should satisfy a perimeter law
discussed at the beginning of this section.

In the spin background, there is no density fluctuation due to
the no-double-occupancy constraint, and locally an increase of $\uparrow$
spinons is always compensated by a decrease of $\downarrow$ spinons. Thus the
distribution of $\uparrow$ spinons will actually reflect  that of all spins
in the hole-absent region. With the
presence of a short-range AF correlation (the  correlation length $\xi> a$),
excess and deficit of $\uparrow$ spinons are neighboring to each other
(a $\xi \sim a$ case is shown in Fig.6).  In such a
``Meissner phase'', the net contribution to $\Delta N_c^{\uparrow}$ mainly
comes from $\uparrow$ spinons close to the path $C$, and $\Delta N_c^
{\uparrow}$ should satisfy a perimeter law instead of an area law as
explained before. Correspondingly, the path
$C_p$ should be always near the path $C$ to account for
$\Delta N_c^{\uparrow}$ in terms of Eq. (4.11),
as shown in Fig.7. An average separation $d_p$ of $C_p$ from $C$  may  be
estimated  as follows: $\pi d_p^2\times B_s\sim \mbox{
one flux quanta} = \pi$, which leads to  $d_p\sim
1/\sqrt{B_s}=d_c$ --- the cyclotron length under $B_s$.

So after one replaces ${\bf A}^s_+$ by ${\bar {\bf A}}^s$ in Eq.(4.7),
the fluctuation effect can be taken into account by deforming
the path of $p$ species from $C$ to $C_p$. The deviation of $C_p$ from $C$
depends on the detailed fluctuation of ${\bf {A}}_+^s$.
But if temperature is sufficiently higher
than the spinon characteristic energy scale (the broadening of the Landau
level in the spinon spectrum), the path  $C_p$ shown in Fig.7
may  be reasonably regarded as  a random one,  with an average separation $d_c$
from the
path $C$. This effectively means that
the binding constraints of $h$, $p$ and $q$ implemented by the Lagrangian
fields ${\mbox {\boldmath $\lambda$ }}$ and ${\mbox {\boldmath $\beta $}}$
have been relaxed within a scale $\sim d_c$. Here the kinetic energy cost of
the deformation path $C_p$ is neglected  due to the smallness of $d_c$. In the
weak gauge-fluctuation case, $d_c$ could be too large  for the present
approximation to be valid.

Thus, the motion of a holon under the influence
of the gauge field ${\bf A}^s$ can be effectively described as a bare holon
bound to a pair of auxiliary species, $p$ and $q$, which are undergoing
cyclotron motions in opposite directions.
These $p$ and $q$ species reflect a ``localized'' effect caused by the phase
interference at short-distances. As $p$ and $q$ are confined in the
Landau levels,
a large degeneracy is involved here. In real space, such a renormalized
holon would look like a polaron, and behave like a diffusive particle in
the absence of external fields.  This peculiar structure will decide a unique
transport phenomenon.

The generalization of the above scheme to
a many-body case is straightforward. Here one has to be cautious about the
statistics of each species in the many-body case. A holon as a composite
particles of $h, p$, and $q$ species has to satisfy the ( hard-core) bosonic
statistics. Thus a symmetric choice would be that the $h$ species corresponds
to a boson while the $p$ and $q$ species are spinless fermions (so that the
hard-core condition can be automatically realized).  Then the
effective many-body Lagrangian as a generalization of the one-body
approximation discussed above can be written down in the functional-integral
formalism as
\begin{eqnarray}
{\cal L}_h=\int d^2 {\bf r} \left\{ h^+ \partial_{\tau } h+ p^+ \partial_{
\tau } p + q^+ \partial _{\tau } q+\lambda (p^+p-h^+h)
+\beta (q^+q-h^+h) \right\}\nonumber \\
+\int d^2{\bf r} \left\{ h^+({\bf r}) \frac {(-i\nabla+ {\mbox {\boldmath $
a$}}^{ext})^2
} {2 m_h} h({\bf r})
+p^+({\bf r}) \frac {(-i\nabla -{\bar {\bf A}}^s )^2} {2 m_p} p({\bf r})
+q^+({\bf r})\frac{(-i\nabla +{\bar {\bf A}}^s )^2} {2 m_q} q({\bf r})
\right\}, \end{eqnarray}
in which $\mbox{\boldmath $ a$}^{ext}$ is an external vector potential. And the
fields
$h({\bf r})$, $p({\bf r})$ and $q({\bf r})$ are in the
coherent representation. The Lagrangian multipliers $\lambda ({\bf r})$ and
$\beta({\bf
r})$ enforce the following binding constraint
\begin{equation}
h^+({\bf r}) h({\bf r})=p^+({\bf r}) p({\bf r})=q^+({\bf r}) q({\bf r}),
\end{equation}
in the length scale larger than $d_c$. An ultraviolet momentum cut-off
$\Lambda=1/d_c $ is then implied in $ \lambda $ and $\beta $ fields in the
Lagrangian ${\cal L}_h$.
The masses $m_h, m_p$ and $m_q$ are determined up to that their total
summation
is equal to $m_t$. The value of each individual mass has to be decided beyond
the present approximation, and this uncertainty will not affect the general
conclusions we shall draw from the Lagrangian ${\cal L}_h$.

Therefore, we obtain an effective long-wavelength Lagrangian (4.12) for the
holons after the short-distance phase-interference is carefully treated. This
is a rather unusual Lagrangian because there are three auxiliary species
involved. But it is not derived in its first time here. A similar one
has already been found\cite{fb2} in the {\em Scheme One}
flux-binding state (see Sec. II) through different method. Even though the
approximations
involved
in these two states are different, their origins are quite similar
and it is not of surprise finding their normal states to be
so  close. In fact, the structures of two effective Lagrangians for holons are
identical,
except that $B_s$ is twice larger in Ref.\onlinecite{fb2} and the $h$ field
is a fermionic one there. A larger $B_s$ in Ref.\onlinecite{fb2} is due to the
$\pi$-flux
phase $\phi^0_{ij}$ has been  incorporated into $B_s$ at continuum
limit. These differences do not change the  canonical
behaviors described below.

The effective Lagrangian  ${\cal L}_h$  determines an
anomalous transport phenomenon\cite{fb2}  which amazingly matches the
essential
characteristics of the optimally doped cuprates. We shall outline the main
results in the following. For detailed discussions, one is referred to
Ref.\onlinecite{fb2}. ${\cal L}_h$  can be treated by the standard
gauge-theory approach. Here it is even simpler for lacking the transverse
fields.  It is easy to show that the longitudinal fields $\lambda$ and $\beta$
will enforce the following current constraint among $h$, $p$, and $q$ species:
\begin{equation}
{\bf J}_h^l={\bf J}_p^l={\bf J}_q^l, \end{equation}
where the superscript $l$ implies the longitudinal channel.  This constraint
is consistent with the density constraint [Eq.(4.13)].  It is important to
note that there is no similar current constraint in the transverse channel
due to the absence of the transverse gauge fields.  The total response to an
external electromagnetic field will be related to each species through
Eq.(4.14).  And different combination rules will thus be found in the
longitudinal  and transverse transport channels, which will lead to
distinctive
Hall angle behavior.

The scattering rates of $h$, $p$ and $q$ are decided by their coupling with
the longitudinal gauge fields $\lambda$ and $\beta$ in Eq.(4.12), whose
dynamics are in  turn determined by coupling to $h$, $p$, and $q$ species.
A self-consistent treatment is required here. Due to the peculiar feature
that $p$ and $q$ stay in the Landau level, one finds that the scattering rate
for $p$ and $q$ goes like linear in temperature, i.e.,
$\frac{\hbar}{\tau_s}= 2\kappa k_BT$,
where $\kappa \sim O(1)$ is independent of the coupling constant (the masses)
and only has a weak doping-dependence (and $\frac 1{\tau_s}\propto \omega $ is
also found
when $\hbar \omega>k_BT$).  And for the $h$ species, the scattering rate has a
$T^2$ behavior:
$\frac{\hbar}{\tau_h}\propto  \frac{(k_BT)^2}{t_h}$. In terms of the constraint
Eq.(4.14), the
total longitudinal
resistivity can be determined\cite{fb2} and at $k_BT\ll t_h$ it is dominant by
$1/\tau_s$ so that  $\rho\propto T $. The linear-$T$ resistivity in the
cuprates is
indeed found\cite{opt} to be  related to
a linear-$T$ relaxation rate. Particularly, the coefficient
is roughly around a numerical factor of $2$ for all the optimal
materials.\cite{opt}
This is very interesting feature and is {\em quantitatively} consistent with
the present
theory, where the
coupling-independent coefficient $2\kappa\sim 2$ is determined by  the unique
structure in the scheme.
The transverse resistivity $\rho_{xy}$ can be also obtained, and as noted
before the
distinctive combination rules in the longitudinal and transverse
channels will lead to new consequences. Namely, the  Hall
angle $\Theta$ as defined by $\cot \Theta=\rho/ \rho_{yx}$ is found to be
related to the second  scattering rate $1/\tau_h$:
$\cot \Theta\propto 1/\tau_h \propto T^2$.
Consequently the Hall coefficient $R_H$ follows a $1/T$ behavior.
The involvement of a second scattering rate $\propto T^2$ in the
Hall angle for high-$T_c$ cuprates was first pointed out by Princeton group
[Ref.\onlinecite{angle1,angle2}] based on the analysis of the experimental
data. The effective Lagrangian in Eq.(4.12) provides a microscopic theory for
it in the first  time. A magneto-resistance with $\Delta \rho/\rho \sim
T^{-4}$
dependence has been also
predicted\cite{fb2} in the longitudinal channel, which has been recently
observed\cite{mr} in $YBCO$ (We note that the overall sign is uncertain in the
theory while experimentally it is found to be positive).  Due to the
``localized'' effect of holon, a  strongly doping-dependent thermopower has
been obtained
in the present framework, which
also agrees well with the experimental measurements in the high-$T_c$
cuprates.

The transport properties determined by the effective Lagrangian  ${\cal L}_h$
may be called as  the canonical ones, which well account for the optimally
doped
cuprates.  In the following we briefly discuss the
condition for deviation from such a canonical case in the present theory. The
key assumption involved in deriving Eq.(4.12) from
$\tilde{H}_h$ is the randomness of the flux quanta movement near the path
$C$ shown in Fig.7, which leads to  uncorrelated segments
along the path $C_p$ so that $p$ and $q$ can be effectively treated as
detached from $h$ at a scale of $d_c$. In this case, the detailed spin
dynamics becomes irrelevant.  In the optimal doped cuprates, the
spinon LLL broadening is very sharp (cf. Sec. III) so that the above
condition may be always satisfied at the normal-state temperature.
For the underdoped $YBCO$, however, the LLL broadening is quite large, and
when the thermal energy is less than it the detailed spin dynamics could
get involved in  $C_p$, especially when $T\sim T_c^*$, and spinons begin to
condensate into the energy bottom.  In this case, $C_p$ may
not be treated as a random one with regard to $C$, and the present
approximation
could break down. As a consequence, a deviation from  the canonical
transport behaviors is expected below some temperature scale, which  should
be correlated with $T_c^*$. Experimentally,  such a correlation between the
transport and the ``spin gap'' is indeed found\cite{ito,cooper} in the
underdoped
$YBCO$ system.
Furthermore, at sufficiently small doping, the transport properties described
by Eq.(4.12) itself can also deviate from the canonical behavior as discussed
in Ref.\onlinecite{fb2}.

Finally, we would like to make a comment on the continuum approximation of
$H_h$. At small doping, the continuum approximation can be well justified for
the spinon
Hamiltonian $H_s$ because $A_{ij}^h$ in it is vanishingly small.
In the present holon case, however, the strong fluctuation effect at
short-distance is
very important, even though $<A_{ij}^h>=0$. Thus one would expect the lattice
effect to be involved
even at small doping. So far we have not been able to include such a lattice
effect. Nevertheless,
as the phase interference at short-distances is not a
coherent effect, the role of the lattice effect may not be really crucial.
Furthermore, $p$ and $q$  are  just auxiliary
particles to take care of the phases interference, which can  be
introduced in the continuum space even when lattice is included. A further
study is still
needed.

\section{CONCLUSION}

In this paper, we have obtained the spin-charge separation scheme based on a
saddle-point state of the $t-J$ model. In this saddle-point state, we find a
deconfinement of spin and charge degree of
freedom at finite doping in 2D case, where the transverse
gauge fluctuation as the confinement force is gapped. Such a gap disappears
at half-filling, where spinons are presumably confined to form spin-1
excitations, and a long-range AF order has been recovered. This saddle-point
state has been also shown to reproduce the known asymptotic
spin-spin correlation in one dimension at both half-filling and finite-doping.
Thus, in some important limits where the behaviors of the $t-J$ model are
known, the present state has produced the right results. These constitute
important check for a strongly-correlated model where the conventional
approximation
breaks down.

The most interesting properties have been found at finite doping in 2D case,
when spin and charge become deconfined. In  this regime, the saddle-point state
becomes meaningful due to the suppression of the gauge fluctuation, and spinons
and holons can be appropriately  treated as
separated systems in terms of conventional approaches. We have shown that in
this saddle-point state there still exist some exotic residual couplings
between spin and charge degrees of  freedom,  in spite of the spin-charge
deconfinement. These  couplings are nonlocal in the sense that  spinons can
feel the
existence
of holons nonlocally by seeing some fictitious flux-quanta bound to the latter,
and {\em vice versa}. Different from a usual electron-phonon system where the
coupling is a single {\em interactive}  term, spinons and holons here are
scattered by each other in distinctive forms. These scattering forces lead to
anomalous spin dynamics and transport  properties in the present system. For
example, a sharp AF peak centered at ${\bf Q}_0$ is exhibited in the imaginary
dynamic spin susceptibility with an energy scale much smaller than the
exchange
energy $J$. The width of such an AF peak in ${\bf q}$-space is determined by
doping concentration in a form $\propto \sqrt{\delta}$, which is
temperature-independent but increases with the energy transfer. The NMR spin
relaxation rate of nuclear spin due to coupling  with such a magnetic
fluctuation shows a non-Korringa behavior for planar $Cu$ nuclei and a strong
suppression for planar $O$ nuclei. Furthermore, we have found a
characteristic temperature $T_c^*$ below which all these AF anomalies get
suppressed, in resemblance to a ``spin gap'' system if the superconducting
temperature $T_c<T_c^*$ and $T_c<T<T_c^*$. Incommensurate AF fluctuation has
also been found in this system at a {\em higher} energy scale. For the charge
degree of freedom, we have demonstrated that the scattering from the spinon
background leads to a strong phase interference at short-distance for holons,
and an effective long-wavelength Lagrangian is derived. Such a Lagrangian has
been found to give the following canonical transport phenomena:
resistivity $\rho\sim T$ with $\hbar/\tau\simeq 2k_BT$; the Hall angle
$\cot \theta_H\propto 1/\tau_h\propto T^2$; the magneto-resistance
$\Delta \rho/\rho \propto T^{-4}$; and a strong doping-dependence  of the
thermopower.

The magnetic and transport properties of the present spin-charge
separation state share amazing similarities with those found in the
high-$T_c$ cuprates, as discussed in the context of the paper. Based on
the theory,
a consistent picture can be conjecturally formed for the normal-state of the
cuprates. The optimally-doped materials can be defined as $T_c\sim T_c^*$,
where the {\em commensurate} AF fluctuation  energy scale becomes very small,
maybe undetectable in terms of  the present neutron-scattering resolution.
Such a small energy scale is a unique feature of the present state, and as
discussed in the context, NMR and neutron-scattering data in both metallic
$LSCO$ and $YB_2Cu_3O_7$ can be reconciled here. And in the transport
channel, the canonical behaviors are exhibited in agreement with the transport
measurements. A larger commensurate  AF fluctuation energy scale can be
realized at a smaller doping regime, where holons tend to be localized, and the
interlayer coupling in $YBCO$ could further help the broadening of the energy
scale. In this regime, a commensurate AF fluctuation may become  observable in
neutron-scattering like in the underdoped $YBCO$, and pseudo spin gap
behaviors
will also show up in both neutron-scattering and NMR spin relaxation rate
below $T_c^*$ ($>T_c$). In this case, a deviation from the canonical
behaviors are expected in the transport channel below a temperature scale
around $T_c^*$. We note that in the present
theory, there is no direct experimental input except that the underlying $t-J$
model is widely perceived as a simplified description of the copper-oxide
layer in the cuprates. This becomes a compelling fact as so many important
experimental features are  naturally exhibited  in the present state. We
have pointed out in the introduction
that the spin-charge separation is a key  for all these delicate magnetic
and transport anomalies to appear.

Two other important issues remain to be clarified in the present spin-charge
separation scheme. One is about the single-electron properties, particularly
the location of Fermi surface, and the other is
about the superconducting condensation in this framework. Since an electron is
described as composed of two {\em bosonic} holon and spinon, the accompanied
nonlocal phase fields in Eq.(1.6) will play a central role to restore the
fermionic properties of the electron. As outlined in the introduction,
the nonlocal phase fields will be responsible for an electronic Fermi surface
satisfying the Luttinger volume theorem as well as a finite pairing order
parameter
$<c^+_{i\uparrow}c^+_{j\downarrow}>$ (and its symmetry) when both holons and
spinons are
Bose condensed. We shall discuss these important problems in the
follow-up papers.

\acknowledgments

The authors acknowledge helpful discussions with T.K. Lee, D. Frenkel, and A.
Chubukov.
Z.Y.W. would like to thank G. Aeppli, G. Baskaran,  B. Doucot, M.
Imada, S. Liang, Y. Ren, A. Sokol, O. Starykh, Z.B. Su,  C. Varma, R.E.
Walstedt,
and L. Yu
for stimulating conversations. The present work is supported partially by
Texas Advanced Research Program
under Grant No. 3652182, a grant from the Robert Welch foundation, and
by Texas
Centre for Superconductivity at University of Houston.

\newpage
\begin{center}
CAPTIONS
\end{center}

Fig. 1. $\chi''({\bf Q}_0, \omega)$ vs. $\omega$. A pseudo spin-gap behavior
is exhibited at $T<T_c^*$,  where the low-energy part ($\simeq 5$$meV$) is
continuously suppressed with the decrease of the temperature.
The spectral function in $\chi''({\bf Q}_0, \omega)$ is chosen to describe $
YBa_2Cu_3O_{6.6}$ (see context).

Fig. 2. The NMR spin-lattice relaxation rate for the planar $Cu$ nuclei
calculated in
terms of the same spectral function used in Fig. 1.

Fig. 3. High-temperature behaviors of the $Cu$ nuclear spin relaxation rate at
various
doping concentration. Notice that the spin relaxation rates are leveled
off at high temperature and  essentially
indistinguishable at small doping ($\delta < 0.2$).

Fig. 4. Typical $\chi''({\bf q}, \omega)$ vs. ${\bf q}$ at various $\omega$'s.
With the increase of $\omega$, the Gaussian form first gets broadened (dotted
curves with $\omega=0.10$ and $0.15$ [$\omega_0$ as the unit]), and then an
incommensurate split occurs when the inter-level transition becomes dominant
(solid curves with $\omega=0.18 $,
$0.20 $ and $0.30 $).

Fig. 5. Fictitious flux-tubes attached to the spinons (solid circles) which are
seen by
the holon (open circle) on a 2D plane.

Fig. 6. An extreme case of short-range AF correlation. The $+$ and $-$ signs
represent a spin configuration, or equivalently, an array of fictitious
flux-tubes
bound to the spinons. The solid closed path cuts through those dimer pairs of
spins and the total flux enclosed satisfies a perimeter law (see Ref.
\onlinecite{gavazzi}).

Fig. 7. An arbitrary closed path $C$ of a holon (solid one) and a corresponding
path $C_p$ for $p$ species (dashed one).


\begin{references}
\bibitem{note2} As usual, $\exp (i A_{ij}^s)$ in Eq.(1.3) will be expanded up
to the quadratic term of $A_{ij}^s$. The leading term in $\tilde{H}_h$ will
be a tight-binding model with the presence of $\pi$-flux per plaquette. A
continuum
expansion near the band bottom will yield an additional $\sqrt{2}$
mass enhancement due to the flux $\phi_{ij}^0$ in $H_h$.
\bibitem{note1} It is interesting to note  that a
characteristic short-distance scale  under  a uniform
magnetic field is the  cyclotron length which can be defined as  the radius
of a circle which encloses a flux just equal to $\pi$: $\pi \times  (\mbox{
cyclotron length})^2  \times \mbox {magnetic field} = \pi$. Beyond the
cyclotron length, phase destruction could  greatly suppress the amplitude of
the wavepacket.
\end{references}
\end{document}